\documentclass[superscriptaddress,aps,prb,twocolumn,groupaddress]{revtex4}
\usepackage{amssymb}
\usepackage{latexsym}
\usepackage{amsmath}
\usepackage[mathcal]{eucal}
\usepackage{graphicx}
\usepackage{dcolumn}
\usepackage{amsfonts}
\usepackage{bm}
\usepackage{epsfig}
\usepackage{array}

\newcommand{\be}{\begin{equation}}
\newcommand{\ee}{\end{equation}}
\newcommand{\bea}{\begin{eqnarray}}
\newcommand{\eea}{\end{eqnarray}}
\newcommand{\bal}{\begin{align}}
\newcommand{\eal}{\end{align}}

\bibpunct{[}{]}{,\!}{n}{,}{,} 

\begin{document}
\title{Orbital order from the on-site orbital attraction}
\author{M. Khodas}
\affiliation{Department of Physics and Astronomy, University of Iowa, Iowa City, Iowa 52242, USA}
\affiliation{Racah Institute of Physics, Hebrew University of Jerusalem, Jerusalem 91904, Israel}
\author{A. V. Chubukov}
\affiliation{School of Physics and Astronomy, University of Minnesota, Minneapolis, MN 55455, USA}

\begin{abstract}
We study the model of Fe-based superconductors with intra-orbital attraction,
designed to favor a spontaneous orbital polarization.
Previous studies of this model within the two orbital approximation indicated that  the leading instability is towards s-wave superconductivity and the subleading one is towards anti-ferro-orbital order, which  breaks the translational symmetry of the crystal. The two-orbital approximation is, however, not consistent with the Fermi surface geometry of Fe superconductors,  as it yields wrong position of one of hole pockets. Here we analyze  the model with the same interaction but with realistic  Fermi surface geometry (two hole-pockets at the centre of the Brillouin Zone and two electron pockets at it's boundary).  We apply the parquet renormalization group (pRG) technique to detect the leading instability upon the lowering of the temperature. We argue  that the pRG analysis strongly favors a $q=0$ orbital order, which in the band basis is a d-wave Pomeranchuk order.
 \end{abstract}

\maketitle

\section{Introduction}

The analysis of the  competition and the interplay between different types of
 electronic order
  remains the key research direction in the studies of Fe-based superconductors (FeSCs)
\cite{Johnston2010,kotliar,Dai2012,C_Annual_Rev_2012,fcs2014}.
The three
 experimentally
 observed macroscopic orders in FeSCs  are magnetism, superconductivity and nematic order.
The nematic order is less conventional than the other two, but it is ubiquitous in all known families of FeSCs and
has been actively investigated
 in the last few years
  \cite{FS2012,fcs2014,Lederer2015,Yamakawa2015,i_paul}.

The nematic order  breaks lattice rotational  $C_4$ symmetry down to $C_2$ and gives rise
to unequal population of $d_{xz}$ and $d_{yz}$ Fe orbitals and to
the anisotropy of magnetic susceptibilities $\chi_{xx} \neq \chi_{yy}$,
, without breaking the spin rotational and time reversal symmetry.
 The imbalance in the orbital population may or may not
 be accompanied by the breaking of the
 translational invariance of the crystal
 ($q=0$ orbital order or finite $q$ orbital order, respectively).
 In the band basis, a $q=0$  order is a zero momentum $d$-wave order in the particle-hole channel.

The origin of the nematic order remains the subject of debates.
In many FeSCs it is likely associated with partial melting of stripe magnetism
 \cite{Chandra1990,Fang2008,Xu2008,Qi2009,Fernandes2010,preemptive2012,Yamase2015,Fanfarillo2015}.
In FeSe, however, the nematic order is not followed by a stripe magnetic order and may be the result of a spontaneous symmetry breaking  between $d_{xz}$ and $d_{yz}$ orbitals
 \cite{efremov,Kruger2009,WeiKu2009,Chen2010,Lv2010,Onari2012}. Interestingly, the nematic order in FeSe emerges at $T_n =85K$, well above superconducting $T_c \sim 8K$.
 To clarify the role of the orbital degrees of freedom in nematicity  and interplay between nematicity and superconductivity  and other potential orders,  it is useful to study the models in which a spontaneous  orbital order is explicitly favoured by the interaction.
The simplest model of this kind is a two-orbital ($d_{xz}/d_{yz}$) model with on-site intra-orbital attraction and inter-orbital repulsion, tailored to favour non-equal
 density of fermions on $d_{xz}$ and $d_{yz}$ orbitals.  The tight-binding fermionic dispersion in this model is obtained from the full 5-orbital tight-binding dispersion by keeping
  only $d_{xz}$ and $d_{yz}$ orbitals.  The two-orbital model has  been  studied within RPA (Refs. \cite{Yamase2013,Yamase2013a}), by weak coupling logarithmical perturbation theory~\cite{ Andrey_2016},
 and by  Quantum Monte Carlo  (Ref.~\cite{ashwin}).
 The outcome is that the two leading instabilities are the ordinary $s-$wave superconductivity and the $q=(\pi,\pi)$ orbital order. At weak coupling, superconductivity wins. At larger couplings, the $(\pi,\pi)$ orbital order may develop first.
  The two-orbital model is, not, however, directly applicable to FeSCs because it places one of hole Fermi surfaces in the wrong place in the Brillouin zone (BZ) --  at $(\pi,\pi)$ instead of $(0,0)$ 1FeBZ.

In this paper we consider the model with the same interaction, but with more realistic band structure with two hole pockets centered at $(0,0)$ and two electron pockets centered at
$(\pi,0)$ and $(0,\pi)$ in the 1FeBZ.
The goal of our study is to analyze the interplay between superconductivity (SC) and $q=0$ orbital order,  and also spin-density wave (SDW), and  charge-density-wave   (CDW) orders.
Several groups have argued~ \cite{Platt2009,Chubukov2008,PhysicaC2009,Maiti2010,Platt2013} that to adequately describe the
interplay between different ordering tendencies one has to include into consideration  the orbital composition of the low-energy excitations and  analyze
   how different interaction channels affect each other.
To do this, we apply the parquet renormalization group technique (pRG).
 This technique is adequate for FeSCs because
the
interactions between fermions with intermediate energies $ W  \gg E \gg E_F$, where $W$ is of order of bandwidth,  are logarithmical not only in the particle-particle
(Cooper) channel, but also in
the particle-hole channel at momenta $(\pi,0)$ and $(0,\pi)$,  due to opposite  signs of the dispersions near hole and electron pockets.  Because the distance between hole and electron pockets
 in momentum space is  a half of the reciprocal lattice vector, a composite effect of two particle-hole excitations gives rise to a logarithmic enhancement of the interaction
 also
  in the $q=0$ Pomeranchuk channel.
   In the situation when renormalizations of the interactions in more than one  channel are logarithmical, the most
  log-divergent
   Feynman graphs are known as parquet diagrams.
The solution of the pRG equations amounts to the summation of all such diagrams.  Physically, pRG equations show how different couplings and susceptibilities
 in various channels evolve as one progressively integrates out high-energy fluctuations.  In all cases studied, the susceptibilities in several channels increase under pRG
  and diverge at some  RG scale
 $L = \log W/E$, where $E$ is the running energy.  The instability develops in the channel in which the susceptibility diverges at the highest energy (i.e., the smallest $L = L_0 = \log W/E_0$). The instability temperature is of order $E_0$.  If susceptibilities  in several channels diverge at the same $L=L_0$,  the most likely outcome is that the order develops in the channel whose susceptibility diverges with the largest exponent.  This reasoning works when $E_0 > E_F$, i.e.,  when  the instability develops before the scale of $E_F$ is reached.  Below $E=E_F$, different channels effectively decouple. Hence, if $E_0 < E_F$,  one should run pRG down to $E = E_F$, obtain the  values of the couplings at this scale, and then independently consider different channels (say, within RPA) using the couplings at $E=E_F$ as the  "bare" couplings.

In our previous work~\cite{CKF2016},
 the two of us and R. Fernandes applied pRG technique to
   the 4-pocket model with repulsive intra-pocket and inter-pocket interactions.
We have found that at intermediate energies the largest susceptibility is in the SDW channel, the one in $s^{+-}$  SC  channel is subleading, and
the
 susceptibility in the orbital order channel is much smaller than the other two.  However at smaller energies SDW and SC channels strongly compete with each other. The
 SC susceptibility eventually gets larger than the one in SDW channel and diverges at RG scale $L=L_0$ as  $\chi_{SC} \propto (L_0 -L)^{-\alpha_{sc}}$, where
 $L = \log W/E$ and $E$ is the running energy.
   However, due to competition with SDW,  the exponent $\alpha_{sc}$ is smaller than its would be BCS value.
    This reduction of the exponent opens up the door for the ``secondary''  channels, like the orbital order channel (the d-wave Pomeranchuk channel in the band basis),
     which also becomes attractive due to a push from spin fluctuations, but does not get weakened due to competition with SDW.  The susceptibility in the d-wave Pomeranchuk channel $\chi_P$  is  smaller than $\chi_{SC}$  at intermediate energies because the bare Pomeranchuk susceptibility is non-logarithmical,
     but may eventually diverge with the exponent $\alpha_{P} > \alpha_{SC}$.
   We found that this is what actually happens. Namely, the d-wave Pomeranchuk susceptibility $\chi_P$ diverges with the exponent $\alpha_P =1$ and becomes the largest near $L=L_0$. As a result, within one-loop pRG, the leading instability upon the lowering of $T$ is towards a spontaneous orbital order.
 This scenario is a plausible one for FeSe~\cite{CKF2016}, however,  it cannot be rigorously justified for the 4-pocket model
  because there $\alpha_{SC}$ is not particularly small,
  and $\chi_P$ becomes larger than $\chi_{SC}$ only in the vicinity of $L_0$,
  where the running couplings are of order one and
  the corrections to one-loop pRG equations are also of order one.

 In this paper we report the results of pRG analysis of the same model as in Ref.\cite{CKF2016} but with intra-pocket attraction. We show that in this model
   s-wave and d-wave SC channels, SDW channel and CDW channel are degenerate and the susceptibilities in all these channels diverge with the same exponent $\chi \propto 1/(L_0 - L)^{\alpha}$.
Because of the competition between many channels, $\alpha$ turns out to be very small: $\alpha = (\sqrt{5} -2)/3 \approx 0.08$. As a result,  these susceptibilities  barely diverge. Meanwhile, the susceptibility in the Pomeranchuk channel still diverges with the exponent $\alpha_P =1$.  Because of the large difference in the values of the exponents, the susceptibility in the Pomeranchuk channel becomes the largest at smaller $L$, where one-loop pRG is under better control.
  In other words,  the fierce competition between the two SC channels, SDW channel, and CDW channel nearly halts the divergencies of the corresponding susceptibilities and allows the Pomeranchuk channel to emerge as a clear winner.
These results differ from the earlier studies in of a two-orbital model where the Pomeranchuk instability was found to be subleading \cite{Andrey_2016,ashwin}.
This can be traced to the competition between the channels described above that is absent in the previously studied model.
The Pomeranchuk instability dominate in the present case since the correlations in other channels are suppressed.

The paper is organized as follows.
In Sec.~\ref{sec:Model} we introduce our model and discuss approximations.
In Sec.~\ref{sec:Bare} we introduce  superconducting, SDW, CDW, and nematic (Pomeranchuk) order parameters
and analyse the development
and hierarchy of different types of order within RPA, i.e. without the inclusion of the couplings between different channels.
In Sec.~\ref{sec:RG}  we include inter-channel couplings and   analyse the flow of the interactions  within pRG.
In Sec.~\ref{sec:fixed} we re-examine the hierarchy of instabilities by evaluating the susceptibilities in different channels along the fixed trajectories of the pRG flow.
We present our conclusions in Sec.~\ref{sec:Discussion}.

\section{The model with intra-orbital attraction}
\label{sec:Model}

The model we study is defined by the Hamiltonian,
\begin{align}\label{H_h}
\mathcal{H} = \mathcal{H}_0 + \mathcal{H}_{int}\, ,
\end{align}
where $\mathcal{H}_0$ is the quadratic part and $\mathcal{H}_{i}$ is the interaction Hamiltonian.
We discuss the effective low-energy band structure model captured by the $\mathcal{H}_0$ first.
We consider 1-Fe BZ with the two hole-pockets at the BZ center and the two electron pockets centered at $\bm{Q}_{1} = (0,\pi)$ and $\bm{Q}_{2} = (\pi,0)$.
Like in Ref. \cite{CKF2016} we treat the  two hole pockets as consisting of $d_{xz}$ and $d_{yz}$ orbitals (as they actually are), and approximate the
 electron pocket at $(0,\pi)$  as consisting of $d_{yz}$ orbital and the one at $(\pi,0)$ as consisting of $d_{xz}$ orbital, i.e., neglect the contributions to
  electron pockets from $d_{xy}$ orbital.

The quadratic part of the Hamiltonian is expressed
 as follows:
\begin{align}
\mathcal{H}_{0}=\sum_{\bm{k},\alpha}\sum_{\mu,\nu=1,2}
\big[
&d_{\mu\alpha}^{\dag}(\bm{k})\mathcal{H}_{\mu,\nu}^{\Gamma}(\bm{k})d_{\nu\alpha}(\bm{k})
\notag \\
& +f_{\mu\alpha}^{\dag}(\bm{k})\mathcal{H}_{\mu,\nu}^{M}(\bm{k})f_{\nu\alpha}(\bm{k})
\big]
\,,\label{Hfree}
\end{align}
where the subscripts $\mu,\nu=1,2$ refer to the $xz$ and $yz$ orbitals respectively, and
\begin{align}
\mathcal{H}^{\Gamma}(\bm{k})\!=\!\begin{bmatrix}\epsilon_{h}\!+\!\frac{k^{2}}{2m_{h}}\!+\!ak^{2}\cos2\theta_{k} & ck^{2}\sin2\theta_{k}\\
ck^{2}\sin2\theta_{k} & \epsilon_{h}\!+\!\frac{k^{2}}{2m_h}\!-\!ak^{2}\cos2\theta_{k}
\end{bmatrix}\label{HGamma}
\end{align}
for states near hole pockets, and
\begin{align}
\mathcal{H}^{M}(\bm{k})\!=\!\begin{bmatrix}\epsilon_{e}\!+\!\frac{k^{2}}{2m_{e}}\!+\!bk^{2}\!\cos2\theta_{k} & 0\\
0 & \epsilon_{e}\!+\!\frac{k^{2}}{2m_{e}}\!-\!bk^{2}\!\cos2\theta_{k}
\end{bmatrix}\label{HM}
\end{align}
for states near electron pockets \cite{Cvetkovic2013}. In Eqs.~\eqref{HGamma} and \eqref{HM}
$\theta_{k}=\arctan(k_{y}/k_{x})$. The parameters $\epsilon_{h,e}$, $1/m_{h,e}$,
$a$, $b$ and $c$  can be either  determined by comparison with the band structure calculations, or, better, taken from experiments.
 To simplify calculations, we set $a=c$ in
Eq.~(\ref{HGamma}), in which case the two hole FSs are circular,
 and the dispersions of the two hole excitations are $\epsilon_h + k^2/(2m_{h1})$ and $\epsilon_h + k^2/(2m_{h2})$, where
 $m_{h1,2} = m_h/(1\pm 2a m_h)$.  To simplify the presentation of pRG results, below we neglect the difference between the two hole masses, i.e., approximate
 $m_{h,1,2} \approx m_h$.  We will also neglect the $b$ term in  Eq.~\eqref{HM}.  We verified that keeping $m_{h1}$ and $m_{h2}$ different complicates the formulas for pRG flow but doesn't affect the results.

We now turn to the interaction Hamiltonian.
 We follow  Refs. \cite{Yamase2013,Yamase2013a,Andrey_2016,ashwin} and consider 4-fermion interaction tailored to favour  a spontaneous orbital polarization:
\begin{align}\label{H_orb}
\mathcal{H}_{int} = - g \sum_j (n_{j,xz} - n_{j,yz})^2\, ,
\end{align}
where the summation index $j$ enumerates the iron sites located at $\bm{R}_{j}$.
The orbital occupation $n_{j,\mu}$ with $\mu = xz,yz$ includes contributions from the
two spin orientations, $n_{j,\mu}=n_{j,\mu\uparrow}+n_{j,\mu\downarrow}$.
For each spin polarization, $\sigma = \uparrow,\downarrow$, the occupation $n_{j,\mu\sigma}=\psi_{j\mu\sigma}^{\dag} \psi_{j\mu\sigma}$, where
\begin{align}
\psi_{j\mu\sigma}=\frac{1}{\sqrt{N}}\sum_{\bm{k}}\left[d_{\mu\sigma}(\bm{k})+f_{\mu\sigma}(\bm{k})e^{i\bm{Q}_{1(2)}\bm{R}_{j}}\right]e^{i\bm{k}\bm{R}_{j}}\label{site_d}
\end{align}
annihilates the electron at the site $\bm{R}_j$ with spin $\sigma$ in the orbital state $\mu$.

The Hamiltonian \eqref{H_orb} is a particular realization of the Hubbard-Hund on-site interaction Hamiltonian,
\begin{align}\label{interaction_K_1}
H_{UJ}=&\frac{U}{2}\sum_{j,\mu}n_{j,\mu}n_{j,\mu}+\frac{U'}{2}\sum_{j,\mu\neq\mu'}n_{j,\mu}n_{j,\mu'}
\notag \\
&+ \frac{J}{2}\sum_{j,\mu'\neq\mu}\sum_{\sigma\sigma'}\psi_{j\mu\sigma}^{\dag}\psi_{j\mu'\sigma'}^{\dag}\psi_{j\mu\sigma'}\psi_{j\mu'\sigma}
\notag \\
&+\frac{J'}{2}\sum_{j,\mu'\neq\mu}\psi_{j\mu\sigma}^{\dag}\psi_{j\mu\sigma'}^{\dag}\psi_{j\mu'\sigma'}\psi_{j\mu'\sigma}\,
\end{align}
with
\begin{align}\label{identify}
U = - 2 g\, , \quad U' = 2 g\, , \quad J = J' = 0\, .
\end{align}
As we discuss below, the actual number of independent interaction constants is higher than one.
In result few interaction channels are degenerate for the model specified by Eq.~\eqref{H_orb}.
For instance, as $J'=0$ the pairing processes for $d_{xz}$- and $d_{yz}$-derived Cooper pairs are independent which makes the $s$ and $d$ wave superconducting pairing degenerate.
We furthermore expect the degeneracy between inter-orbital SDW and CDW channels as in this case the direct processes contribution of Eq.~\eqref{H_orb} are absent regardless of the state of spin polarization of interacting electrons.
These expectations are confirmed by explicit evaluation in Sec.~\ref{sec:Bare}.

The original interaction Hamiltonian has just one coupling $g$ and one may think that one needs just one pRG equation  for the flow of $g$. However, earlier pRG studies of
FeSCs already indicated that this is not the case for two reasons. First, under pRG, $U$ and $U'$ become non-equivalent, and $J$ and $J'$ are generated.
 Second, the full on-site interaction Hamiltonian does not remain invariant under pRG, i.e, new interactions are generated, which can be identified as interactions between fermions at
 neighboring sites.  One can make sure (see Ref.  \cite{CKF2016}  for details) that the total number of different $C_4$-symmetric
 4-fermion combinations of low-energy fermions from Eq.~\eqref{Hfree} is equal to 14.  The corresponding Hamiltonian is
\begin{align}\label{H_int13}
H = \sum_{j =1}^5 H_{U_j}\, ,
\end{align}
where
\begin{align}\label{U1}
H_{U_1}&=  U_{1}\sum\nolimits'\left[f_{1\sigma}^{\dag}f_{1\sigma}d_{1\sigma'}^{\dag}d_{1\sigma'}+f_{2\sigma}^{\dag}f_{2\sigma}d_{2\sigma'}^{\dag}d_{2\sigma'}\right]
\notag \\
+ &\bar{U}_{1}\sum\nolimits'\left[f_{2\sigma}^{\dag}f_{2\sigma}d_{1\sigma'}^{\dag}d_{1\sigma'}+f_{1\sigma}^{\dag}f_{1\sigma}d_{2\sigma'}^{\dag}d_{2\sigma'}\right]
\end{align}
\begin{align}\label{U2}
H_{U_2}&=  U_{2}\sum\nolimits'\left[f_{1\sigma}^{\dag}d_{1\sigma}d_{1\sigma'}^{\dag}f_{1\sigma'}+f_{2\sigma}^{\dag}d_{2\sigma}d_{2\sigma'}^{\dag}f_{2\sigma'}\right]
\notag\\
+ & \bar{U}_{2}\sum\nolimits'\left[f_{1\sigma}^{\dag}d_{2\sigma}d_{2\sigma'}^{\dag}f_{1\sigma'}+f_{2\sigma}^{\dag}d_{1\sigma}d_{1\sigma'}^{\dag}f_{2\sigma'}\right]
\end{align}
\begin{align}\label{U3}
H_{U_3}&=  \frac{U_{3}}{2}\sum\nolimits'\left[f_{1\sigma}^{\dag}d_{1\sigma}f_{1\sigma'}^{\dag}d_{1\sigma'}+f_{2\sigma}^{\dag}d_{2\sigma}f_{2\sigma'}^{\dag}d_{2\sigma'}+h.c.\right]
\notag \\
&+\frac{\bar{U}_{3}}{2}\sum\nolimits'\left[f_{1\sigma}^{\dag}d_{2\sigma}f_{1\sigma'}^{\dag}d_{2\sigma'}+f_{2\sigma}^{\dag}d_{1\sigma}f_{2\sigma'}^{\dag}d_{1\sigma'}+h.c.\right]
\end{align}
\begin{align}\label{U4}
H_{U_4}&=   \frac{U_{4}}{2}\sum\nolimits'\left[d_{1\sigma}^{\dag}d_{1\sigma}d_{1\sigma'}^{\dag}d_{1\sigma'}+d_{2\sigma}^{\dag}d_{2\sigma}d_{2\sigma'}^{\dag}d_{2\sigma'}\right]
\notag \\
+& \frac{\bar{U}_{4}}{2}\sum\nolimits'\left[d_{1\sigma}^{\dag}d_{2\sigma}d_{1\sigma'}^{\dag}d_{2\sigma'}+d_{2\sigma}^{\dag}d_{1\sigma}d_{2\sigma'}^{\dag}d_{1\sigma'}\right]\notag\\
 +& \tilde{U}_{4}\sum\nolimits'd_{1\sigma}^{\dag}d_{1\sigma}d_{2\sigma'}^{\dag}d_{2\sigma'}+\tilde{\tilde{U}}_{4}\sum\nolimits'd_{1\sigma}^{\dag}d_{2\sigma}d_{2\sigma'}^{\dag}d_{1\sigma'}
\end{align}
\begin{align}\label{U5}
H_{U_5}&=
\frac{U_{5}}{2}\sum\nolimits'\left[f_{1\sigma}^{\dag}f_{1\sigma}f_{1\sigma'}^{\dag}f_{1\sigma'}+f_{2\sigma}^{\dag}f_{2\sigma}f_{2\sigma'}^{\dag}f_{2\sigma'}\right]
\notag \\
+ & \frac{\bar{U}_{5}}{2}\sum\nolimits'\left[f_{1\sigma}^{\dag}f_{2\sigma}f_{1\sigma'}^{\dag}f_{2\sigma'}+f_{2\sigma}^{\dag}f_{1\sigma}f_{2\sigma'}^{\dag}f_{1\sigma'}\right]\notag\\
+ & \tilde{U}_{5}\sum\nolimits'f_{1\sigma}^{\dag}f_{1\sigma}f_{2\sigma'}^{\dag}f_{2\sigma'}+\tilde{\tilde{U}}_{5}\sum\nolimits'f_{1\sigma}^{\dag}f_{2\sigma}f_{2\sigma'}^{\dag}f_{1\sigma'}\, .
\end{align}
In Eqs.~(\ref{U1}-\ref{U5}) the notation $\sum\nolimits'$ stands for the summation over spins and over the momenta subject to the momentum conservation.
For instance,
\begin{align}\label{define_sum'}
\sum\nolimits'  f_{1\sigma}^{\dag}f_{1\sigma}d_{1\sigma'}^{\dag}d_{1\sigma'} & = \notag \\
  = \sum_{\bm{k}_1,\bm{k}_2,\bm{k}_3,\bm{k}_4} \sum_{\sigma,\sigma'} & f_{1\sigma}^{\dag}(\bm{k}_1)f_{1\sigma}(\bm{k}_2)d_{1\sigma'}^{\dag}(\bm{k}_3)d_{1\sigma'}(\bm{k}_4)
\notag \\
& \times \delta_{\bm{k}_1+\bm{k}_2 +\bm{k}_3 + \bm{k}_4,0} \, ,
\end{align}
where $\delta$ in the last line stands for the Kronecker $\delta$.

At the bare level
\begin{align}
U_1 &= U_2 = U_3 = U_4 = U_5 = - 2g
,\notag\label{Hubbard_relation_1}\\
\bar{U}_{1} & =\tilde{U}_{4}=\tilde{U}_{5}= 2 g
\end{align}
and other  interactions are zero. But all 14 interactions are generally generated under pRG, i.e., the full set of pRG equations contains 14 coupled equations.
One can easily make sure that no other terms are generated by pRG.

Because pRG calculations involve fermions near hole and electron pockets, it is advantageous to move to the band basis, i.e.,  diagonalize the quadratic Hamiltonian
 for excitations near hole pockets and re-express the interaction Hamiltonian in terms of band operators.  We refrain from  presenting the
  corresponding Hamiltonian  as the formula for it is quite lengthy.

\section{Order parameters and susceptibilities within RPA }
\label{sec:Bare}

We begin the discussion of potential ordered states in the model of Eq.~\eqref{H_orb} by
 first treating all channels as independent and analysing the corresponding susceptibilities within RPA.
   In order to avoid complex formulas, we  present the order parameters in the orbital basis and
 list the results of the computations of the susceptibilities within RPA.
   The actual computations of the susceptibilities were performed in the band basis.

\subsection{SDW channels}

There are two SDW orders with momenta $(0,\pi)$ and $(\pi,0)$. One involves bilinear
combinations of fermions from the same orbital, the other involves fermions from different orbitals.

The two intra-orbital SDW order parameters are constructed of $f^\dagger_\alpha d_\alpha$, which are diagonal in the orbital index:
\begin{align}
{\bm{s}}_{1,2}^{r}& =f_{1,2}^{\dag}\bm{\sigma}d_{1,2} +d_{1,2}^{\dag}\bm{\sigma}f_{1,2},
\notag \\
{\bm{s}}_{1,2}^{i}&=i(f_{1,2}^{\dag}\bm{\sigma}d_{1,2}- d_{1,2}^{\dag}\bm{\sigma}f_{1,2})\,.\label{s1}
\end{align}
We will refer to ${\bm{s}}_{1,2}^{r}$ and ${\bm{s}}_{1,2}^{i}$ as to real and imaginary SDW order parameters.
The real ${\bm{s}}_{1,2}^r$  gives rise to a SDW on Fe cites, and  ${\bm{s}}_{1,2}^i$ gives rise to a spin current.

The  inter-orbital anti-ferromagnetism is described by the order parameters
\begin{align}
\bar{\bm{s}}_{1,2}^{r}&=f_{1,2}^{\dag}\bm{\sigma}d_{2,1} +d_{1,2}^{\dag}\bm{\sigma}f_{2,1},
\notag \\
\bar{\bm{s}}_{1,2}^{i}&=i(f_{1,2}^{\dag}\bm{\sigma}d_{2,1}- d_{1,2}^{\dag}\bm{\sigma}f_{2,1})\, \label{s1_b}
\end{align}
which are off diagonal in the orbital index.
The real ${\bar {\bm{s}}}_{1,2}^r$  gives rise to an unconventional SDW, which in real space is concentrated on pnictogen/chalcogen sites and has no weight on Fe sites, and
${\bar {\bm{s}}}_{1,2}^r$  gives rise to a corresponding spin current.

\begin{figure}
\includegraphics[width=0.8\columnwidth]{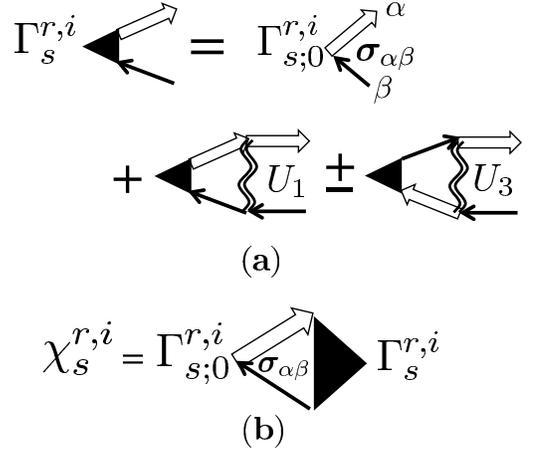}
\caption{
(a) Graphical representation of the Dyson equation for the vertices $\Gamma_{\bm{s}}$ in the SDW channel.
The equation \eqref{n_n1}  for $\bar{\Gamma}_{\bm{s}}$ is obtained by replacing all $U_i$ by ${\bar U}_i$.
Single (double) lines represent propagators of fermions near hole (electron) pockets.
(b) Graphical representation of the RPA formula for the susceptibility in the spin channel, Eq.~\eqref{eq:chi}.
\label{fig:G_s}}
\end{figure}

The part of the interaction Hamiltonian \eqref{H_int13}, bilinear in SDW order parameter, is
\begin{align}\label{H_s}
H_{\bm{s},\pi}& =  \frac{1}{8}(-U_{1}-U_{3})\left[\bm{s}_{1}^{r}\bm{s}_{1}^{r}+\bm{s}_{2}^{r}\bm{s}_{2}^{r}\right]
\notag \\
+ &\frac{1}{8}(-U_{1}+U_{3})\left[\bm{s}_{1}^{i}\bm{s}_{1}^{i}+\bm{s}_{2}^{i}\bm{s}_{2}^{i}\right]\notag\\
 + &\frac{1}{8}(-\bar{U}_{1}-\bar{U}_{3})\left[\bar{\bm{s}}_{1}^{r}\bar{\bm{s}}_{1}^{r}+\bar{\bm{s}}_{2}^{r}\bar{\bm{s}}_{2}^{r}\right]
 \notag \\
 +&\frac{1}{8}(-\bar{U}_{1}+\bar{U}_{3})\left[\bar{\bm{s}}_{1}^{i}\bar{\bm{s}}_{1}^{i}+\bar{\bm{s}}_{2}^{i}\bar{\bm{s}}_{2}^{i}\right]\,.
\end{align}
At the bare level, the interaction between $\bm{s}_{1,2}^{r}$ is repulsive: $ (- U_1 - U_3)/8 =g/2>0$, while the interaction between $\bm{s}_{1,2}^{i}$ vanishes.
The interactions between $\bar{\bm{s}}_{1,2}^{r}$ and between $\bar{\bm{s}}_{1,2}^{i}$ are attractive and have the same magnitude: $ (-\bar{U}_1 \mp \bar{U}_3)/8 = -g/4 < 0$.
Adding the terms $ {\bar \Gamma}_{s;0} {\bm s}_{1,2}^{r,i}$  with infinitesimally small prefactors ${\bar \Gamma}_{s;0}$ to the Hamiltonian
and summing up ladder series of renormalizations of ${\bar \Gamma}$ we obtain (see Fig.~\ref{fig:G_s}a)
\begin{equation}
{\bar \Gamma}_{s;1,2}^{r,i} = \frac{{\bar \Gamma}_{s;0}}{1 - 2 {\bar g}\log{W/T}}\, ,
\label{n_n1}
\end{equation}
where we introduced
\begin{align}
\bar{g} = \frac{ g m }{ 4 \pi}
\end{align}
with $m = 2 m_h m_e/(m_h + m_e)$, and $m_h$ and $m_e$ are the masses for excitations
 near hole and electron pockets, see Eqs. \eqref{HGamma} and ~\eqref{HM} and the discussion after them.

 Eq.~\eqref{n_n1} holds for
$T$ larger than a typical energy below which the logarithm in the particle-hole channel at momenta $(0,\pi)$ and $(\pi,0)$  is cut (Ref. \cite{vvc}).
 We see that, within RPA, inter-orbital magnetic instability develops at the temperature $T_{sdw}$ at which
\begin{align}\label{inst_13}
2 \bar{g} \log\frac{W}{T_{sdw}} = 1\, .
\end{align}
The same result can be obtained by analyzing the susceptibilities within RPA. The  bare susceptibilities in ${\bm s}_{1,2}^{r,i}$ channels are
$\chi_0 (T) = (2 m/\pi) \log{W/T}$.    Within RPA,
 the full susceptibilities in ${\bm s}_{1,2}^{r,i}$ channels are (see Fig.~\ref{fig:G_s}b),
\begin{equation}\label{eq:chi}
{\bar \chi}_{s;1,2}^{r,i} (T) = \frac{\chi_0 (T)}{1 - (g/4) \chi_0 (T)} = \frac{\chi_0 (T)}{1 - 2 {\bar g}\log{W/T}}\, .
\end{equation}
The susceptibilities obviously diverge at the same $T_{sdw}$  as the vertices ${\bar \Gamma}_{1,2}^{r,i}$.

\subsection{CDW channels}

\begin{figure}
\includegraphics[width=0.8\columnwidth]{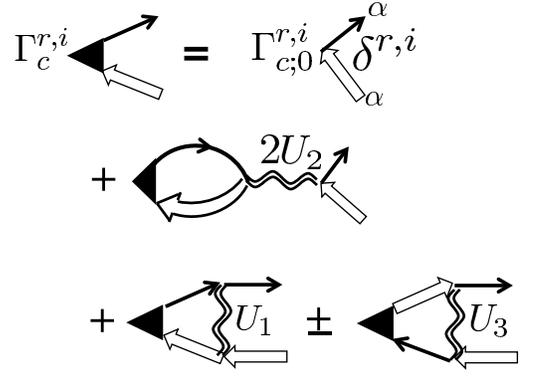}
\caption{
Graphical representation of the Dyson equation for the vertices, $\Gamma_{\bm{c}}$ in the CDW channel.
The infinitesimal external perturbation giving rise to the vertex is proportional to $\delta^{r,i}$ introduced in Eq.~\eqref{del_r} which is diagonal in spin indices.
The equation \eqref{n_n1_1} for $\bar{\Gamma}_{\bm{s}}$ is obtained by replacing all $U_i$ by ${\bar U}_i$ and $\delta^{r,i}$ by $\bar{\delta}^{r,i}$.
\label{fig:G_c}}
\end{figure}

We next consider CDW order parameters with momenta $(\pi,0)$ and $(0,\pi)$.
Like in SDW case, we have two types of order parameters: diagonal and non-diagonal in orbital index.
The order parameters diagonal in the orbital index are
\begin{align}
\delta_{1,2}^{r}&=f_{1,2}^{\dag}d_{1,2}+d_{1,2}^{\dag}f_{1,2},
\notag \\
\delta_{1,2}^{i}&=i(f_{1,2}^{\dag}d_{1,2}- d_{1,2}^{\dag}f_{1,2})
\, ,\label{del_r}
\end{align}
and the ones non-diagonal in the orbital index are
\begin{align}
\bar{\delta}_{1,2}^{r}& =f_{1,2}^{\dag}d_{2,1}+ d_{2,1}^{\dag}f_{1,2},
\notag \\
\bar{\delta}_{1,2}^{i}& =i(f_{1,2}^{\dag}d_{2,1}- d_{2,1}^{\dag}f_{1,2})
\, .\label{bar_del}
\end{align}
The order parameter which gives rise to CDW on Fe cites is $\delta_{1,2}^r$. The order parameter  $\delta_{1,2}^i$ gives rise to charge current.
The corresponding interaction terms, bilinear in $\delta_{1,2}^{r,i}$ and ${\bar \delta}_{1,2}^{r,i}$ are
\begin{align}
H_{\delta,\pi}= & \frac{1}{8}(-U_{1}+2U_{2}+U_{3})\left[\delta_{1}^{r}\delta_{1}^{r}+\delta_{2}^{r}\delta_{2}^{r}\right]
\notag \\
+& \frac{1}{8}(-U_{1}+2U_{2}-U_{3})\left[\delta_{1}^{i}\delta_{1}^{i}+\delta_{2}^{i}\delta_{2}^{i}\right] \notag \\
+&\frac{1}{8}(-\bar{U}_{1}+2\bar{U}_{2}+\bar{U}_{3})\left[\bar{\delta}_{1}^{r}\bar{\delta}_{1}^{r}+\bar{\delta}_{2}^{r}\bar{\delta}_{2}^{r}\right]
 \notag\\
 +&\frac{1}{8}(-\bar{U}_{1}+2\bar{U}_{2}-\bar{U}_{3})\left[\bar{\delta}_{1}^{i}\bar{\delta}_{1}^{i}+\bar{\delta}_{2}^{i}\bar{\delta}_{2}^{i}\right]\, .
\label{H_del}
 \end{align}
Performing the same analysis as in the previous section, i.e., adding to the Hamiltonian the extra terms $\Gamma_{c;0} \delta_{1,2}^{r,i}$ and ${\bar \Gamma}_{c;0}
\bar{\delta}_{1,2}^{r,i}$ with infinitesimally small $\Gamma_{c;0}$ and summing up ladder diagrams for the renormalization of the vertices in $\delta_{1,2}^{r,i}$
 and ${\bar \delta}_{1,2}^{r,i}$ channels, we find two results.   First, the  interaction in both inter-orbital CDW channels is $(-{\bar U}_1 + 2 {\bar U}_2 \pm {\bar U}_3)/8 = -g/4$.
 The equation for the full vertex ${\bar \Gamma}_{c;1,2}^{r,i}$ then has the same form as Eq. (\ref{n_n1}) for the SDW vertex, see Fig.~\ref{fig:G_c}:
 \begin{equation}
{\bar \Gamma}_{c;1,2}^{r,i} = \frac{{\bar \Gamma}_{c;0}}{1 - 2 {\bar g}\log{W/T}}\, .
\label{n_n1_1}
\end{equation}
Accordingly, the instability temperature in this channel is the same as for inter-orbital SDW, see Eq. (\ref{inst_13}).

Second, the interaction in the $\delta_{1,2}^i$ channel vanishes, and the one in $\delta_{1,2}^r$  channel (a conventional CDW channel) is attractive:
 $(-U_1+2U_2 + U_3)/8 =  -g/2$.  Accordingly, the vertex renormalization is given by
  \begin{equation}
\Gamma_{c;1,2}^{r} =\frac{\Gamma_{c;0}}{1 - 4 {\bar g}\log{W/T}}\, .
\label{n_n1_2}
\end{equation}
The coupling in (\ref{n_n1_2}) is twice larger than in (\ref{n_n1_1}), hence  the leading instability in the CDW subset is towards a conventional CDW order
$\delta_{1,2}^r$. The corresponding instability temperature $T_{cdw}$ is the solution of
\begin{align}\label{inst_13_1}
4 \bar{g} \log\frac{W}{T_{cdw}} = 1\, .
\end{align}

\subsection{Superconducting channels}

We now turn to the Cooper channel.
We introduce
\begin{align}
\kappa_{\mu\mu'}^{f}=f_{\mu\uparrow}f_{\mu'\downarrow}
\,,\quad\kappa_{\mu\mu'}^{d}=d_{\mu\uparrow}d_{\mu'\downarrow}\,.
\label{Co}
\end{align}
and classify fermion bilinear operators with zero total momentum via the one-dimensional irreducible representations of the $D_{4h}$ point group $A_{1g}$, $B_{1g}, B_{2g}$ and $A_{2g}$
as
\begin{align}
\kappa_{A_{1g}}^{f(d)} & =\kappa_{11}^{f(d)}+\kappa_{22}^{f(d)}\notag\label{C_channels}\\
\kappa_{B_{1g}}^{f(d)} & =\kappa_{11}^{f(d)}-\kappa_{22}^{f(d)}\notag\\
\kappa_{B_{2g}}^{f(d)}&=\kappa_{12}^{f(d)}+\kappa_{21}^{f(d)} \notag\\
\kappa_{A_{2g}}^{f(d)}&=\kappa_{12}^{f(d)}-\kappa_{21}^{f(d)}.
\end{align}
The subscript $g$ in the labels implies that the order parameters are even under  inversion.
The $A_{2g}$ combination
vanishes for a singlet pairing because it is odd in the orbital indices.

The interaction terms bilinear in $\kappa$ are  obtained from Eq.~\eqref{H_int13} by setting
the momenta of the two creation operators appearing in each separate term in Eqs.~(\ref{U1}--\ref{U5}) to be opposite.
(See Eq.~\eqref{define_sum'} for the explicit definition of these terms).
The resulting interaction decouples between different symmetries:
\begin{align}\label{Hkappa}
H_{\kappa}=H_{\kappa_{A_{1}}}+H_{\kappa_{B_{1}}}+H_{\kappa_{B_{2}}}\, .
\end{align}
Each term in Eq.~\eqref{Hkappa} is expressed  through the bilinear components, Eq.~\eqref{C_channels} as
\begin{align}\label{s_SC}
H_{\kappa_{A_{1}}}& =\frac{1}{2}(U_{5}+\bar{U}_{5})[\kappa_{A_{1}}^{f}]^{\dag}\kappa_{A_{1}}^{f}+\frac{1}{2}(U_{4}+\bar{U}_{4})[\kappa_{A_{1}}^{d}]^{\dag}\kappa_{A_{1}}^{d}
\notag\\
+ & \frac{1}{2}(U_{3}+\bar{U}_{3})([\kappa_{A_{1}}^{f}]^{\dag}\kappa_{A_{1}}^{d}+h.c.)\, ,
\end{align}
\begin{align}\label{d_SC}
H_{\kappa_{B_{1}}}& =\frac{1}{2}(U_{5}-\bar{U}_{5})[\kappa_{B_{1}}^{f}]^{\dag}\kappa_{B_{1}}^{f}+\frac{1}{2}(U_{4}-\bar{U}_{4})[\kappa_{B_{1}}^{d}]^{\dag}\kappa_{B_{1}}^{d}
\notag \\
+ & \frac{1}{2}(U_{3}-\bar{U}_{3})([\kappa_{B_{1}}^{f}]^{\dag}\kappa_{B_{1}}^{d}+h.c.)\, ,
\end{align}
\begin{align}\label{HC}
H_{\kappa_{B_{2}}}=\frac{1}{2}(\tilde{U}_{5}+\tilde{\tilde{U}}_{5})[\kappa_{B_{2}}^{f}]^{\dag}\kappa_{B_{2}}^{f}+\frac{1}{2}(\tilde{U}_{4}+\tilde{\tilde{U}}_{4})[\kappa_{B_{2}}^{d}]^{\dag}\kappa_{B_{2}}^{d}\, .
\end{align}
In the $B_{2}$ channel represented by Eq.~\eqref{HC} the interactions involving fermions near hole and electron pockets decouple, and the interactions are repulsive:
$ \tilde{U}_{5}+\tilde{\tilde{U}}_{5} = \tilde{U}_{4}+\tilde{\tilde{U}}_{4} = 2g$.
 As a result, there is no SC instability in the $B_{2g}$ channel within RPA.

\begin{figure}
\includegraphics[width=1.0\columnwidth]{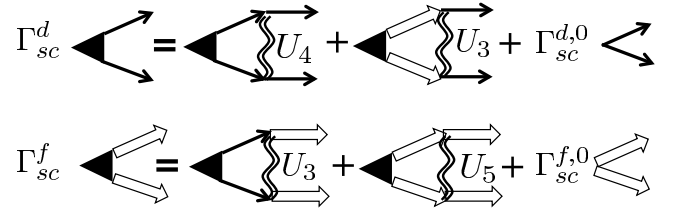}
\caption{
Graphical representation of the Dyson equation \eqref{dG_SC7} for the vertices $\Gamma_{\bm{sc}}^{f,d}$ in spin-singlet $A_{1g}$ ad $B_{1g}$ Cooper channels.
Antisymmetrization with respect to spin indices is assumed.  The equations for $A_{1g}$ and $B_{1g}$ are the same, but the bare vertices
$\Gamma^0_{sc;A_{1g}(B_{1g}}$ and $\Gamma^0_{sc;A_{1g}(B_{1g}}$ have different
 symmetry properties (both are labeled as $\Gamma^0_{sc}$ on the figure).
\label{fig:SC_RPA}}
\end{figure}

In $A_{1g}$ ($s$-wave) and $B_{1g}$ ($d$-wave) Cooper channels  we have, at the bare level,  $U_{5}+\bar{U}_{5} = U_{5}- \bar{U}_{5} = -2g$, $U_{4}+\bar{U}_{4} = U_{4}- \bar{U}_{4}
 = -2g$, $U_{3}+\bar{U}_{3} = U_{3}- \bar{U}_{3} = -2g$.  Comparing (\ref{s_SC}) and (\ref{d_SC}) we immediately find that $A_{1g}$ and $B_{1g}$  channels are degenerate.
Introducing the order parameters $\kappa_{A_{1g}}^{f(d)}$ and $\kappa_{B_{1g}}^{f(d)}$ with the bare vertices $\Gamma_{sc;A_{1g}(B_{1g})}^{f,d} = \Gamma_{sc,0}$
 into the Hamiltonian and summing up series of ladder  renormalizations, see Fig.~\ref{fig:SC_RPA} we obtain
\begin{align}
\Gamma_{sc;A_{1g}(B_{1g})}^{f}& =-\Gamma_{sc;A_{1g}(B_{1g})}^{f}u_{5} L-\Gamma_{sc;A_{1g}(B_{1g})}^{d}u_{3a}L + \Gamma_{sc}^{f,0}
\notag\label{dG_SC7}\\
\Gamma_{sc,A_{1g}(B_{1g})}^{d}& =-\Gamma_{sc;A_{1g}(B_{1g})}^{f} u_{3b}L -\Gamma_{sc;A_{1g}(B_{1g})}^{d} u_{4}L + \Gamma_{sc}^{d,0}\, ,
\end{align}
where $L = \log(W/T)$ is the Cooper logarithm, and the dimensionless interactions are
\begin{align}\label{dimensionless}
u_4 &= U_4 m_h/ (4 \pi), u_5 = U_5 m_e/ (4 \pi),
\notag \\
u_{3a,b} &= U_3 m_{h,e}/ (4 \pi) \, .
\end{align}
Introducing  the matrix
\begin{align}
M_{SC} = - L \begin{bmatrix}
u_5  & u_3  \\
u_3  & u_4
\end{bmatrix}
\end{align}
we can rewrite  Eq. \eqref{dG_SC7} as
\begin{align}
\begin{bmatrix}
\Gamma_{sc;A_{1g},B_{1g}}^{f}\\
\Gamma_{sc;A_{1g},B_{1g}}^{d}
\end{bmatrix}= M_{SC}
\begin{bmatrix}\Gamma_{sc;A_{1g},B_{1g}}^{f}\\
\Gamma_{sc;A_{1g},B_{1g}}^{d}
\end{bmatrix}
+
\begin{bmatrix}
 \Gamma^0_{sc;A_{1g},B_{1g}} \\
  \Gamma^0_{sc;A_{1g},B_{1g}}
 \end{bmatrix}
\, . \label{dG_SC9}
\end{align}
The instability occurs once the largest of the  eigenvalues of the matrix $M_{SC}$ reaches unity.
There are two eigenvalues of $M_{SC}$, equal in $A_{1g}$ and $B_{1g}$ channels. In $A_{1g}$ channel they describe
 the sign preserving s-wave order parameter $s^{++}$ and the order parameter $s^{+-}$, which changes the sign between the hole and electron Fermi surfaces.
 In $B_{1g}$ channel the corresponding eigenvalues  describe a conventional $d-$wave order parameter $(d^{++})$ and
 $d^{+-}$ order parameter which additionally changes the sign between the hole and electron Fermi surfaces.
 Evaluating the eigenvalues we immediately find that $\lambda_{s^{++}} = \lambda_{d^{++}}> \lambda_{s^{+\pm}} = \lambda_{d^{+\pm}}$, hence the first instability upon the lowering
  of $T$ is in the $++$ channel ($s$-wave or $d-$wave). The corresponding eigenvalue is
\begin{align}\label{dG_SC10}
\lambda_{s^{++}} = \lambda_{d^{++}}  & = \left[- (u_4 + u_5) + \sqrt{\left(\frac{u_4-u_5}{2}\right)^2 + u_{3a} u_{3b}} \right]L
\end{align}
Substituting the bare values of the couplings, we find
\begin{align}\label{dG_SC10_1}
\lambda_{s^{++}} = \lambda_{d^{++}}  & = 4{\bar g} L
\end{align}
Note that this result holds for any ratio of the masses $m_e/m_h$.
 The  superconducting $T_c$ in $s^{++}$ and $d^{++}$ channels is then determined from
\begin{align}\label{dG_SC11}
4 \bar{g} \log\frac{W}{T_c} = 1\, .
\end{align}
Comparing with  Eq.~\eqref{inst_13_1}, we find that $T_c$ and $T_{cdw}$ coincide, i.e., within RPA, two superconducting channels and a conventional CDW channel are degenerate in the sense
 that the instability temperatures are the same in all three channels.
Intra-orbital SDW and CDW channels are also degenerate, but the instability temperatures in these channels are smaller.

\subsection{Particle-hole channels at zero momentum transfer}

We next analyze the potential instabilities in the particle-hole channel that do not  break  the translational symmetry of the crystal.
The corresponding order parameters  involve bilinear fermion combinations
\begin{align}
\rho_{\mu\mu'}^{f}=\sum_{\sigma}f_{\mu\sigma}^{\dag}f_{\mu'\sigma}\,,\quad\rho_{\mu\mu'}^{d}=\sum_{\sigma}d_{\mu\sigma}^{\dag}d_{\mu'\sigma}\, ,\label{rho}
\end{align}
Like we did for superconductivity, we
classify fermion bilinear operators with zero transferred momentum
 via the  irreducible representations of the $D_{4h}$ point group.
The combinations in (\ref{rho})  are even under inversion and their transformation   includes only one-dimensional irreducible representations
 $A_{1g}$, $B_{1g}, B_{2g}$ and $A_{2g}$. We omit subscript $g$ below to simplify the notations.

A simple experimentation shows that the combinations of $\rho_{\mu\mu'}^{f,d}$, which transform as a particular representation, are
\begin{align}
\rho_{A_{1}}^{f(d)} & =\rho_{11}^{f(d)}+\rho_{22}^{f(d)}\notag\label{rho_channels}\\
\rho_{B_{1}}^{f(d)} & =\rho_{11}^{f(d)}-\rho_{22}^{f(d)}\notag\\
\rho_{A_{2}}^{f(d)} & =i(\rho_{12}^{f(d)}-\rho_{21}^{f(d)})\notag\\
\rho_{B_{2}}^{f(d)} & =\rho_{12}^{f(d)}+\rho_{21}^{f(d)}\, .
\end{align}

To obtain the interactions in the particle-hole charge channel at
zero momentum transfer   we set $\bm{k}_{1}=\bm{k}_{2}$ or $\bm{k}_{1}=\bm{k}_{4}$
in Eq.~\eqref{H_int13}. Expressing Eq.~\eqref{H_int13} in terms
of the combinations \eqref{rho_channels} we obtain
\begin{align}
H_{\rho}=H_{\rho_{A_{1}}}+H_{\rho_{A_{2}}}+H_{\rho_{B_{1}}}+H_{\rho_{B_{2}}}\,,
\end{align}
where
\begin{align}
H_{\rho_{A_{1}}}& =\frac{1}{8}(U_{5}+2\tilde{U}_{5}-\tilde{\tilde{U}}_{5})[\rho_{A_{1}}^{f}]^{2}+\frac{1}{8}(U_{4}+2\tilde{U}_{4}-\tilde{\tilde{U}}_{4})[\rho_{A_{1}}^{d}]^{2}
\notag \\
& + \frac{1}{4}\rho_{A_{1}}^{f}\rho_{A_{1}}^{d}(2U_{1}-U_{2}+2\bar{U}_{1}-\bar{U}_{2})\label{H_r_A1}
\end{align}
\begin{align}
H_{\rho_{B_{1}}}& =\frac{1}{8}(U_{5}-2\tilde{U}_{5}+\tilde{\tilde{U}}_{5})[\rho_{B_{1}}^{f}]^{2}+\frac{1}{8}(U_{4}-2\tilde{U}_{4}+\tilde{\tilde{U}}_{4})[\rho_{B_{1}}^{d}]^{2}
\notag \\
& +\frac{1}{4}\rho_{B_{1}}^{f}\rho_{B_{1}}^{d}(2U_{1}-U_{2}-2\bar{U}_{1}+\bar{U}_{2})\label{chu_4}
\end{align}
\begin{align}
H_{\rho_{A_{2}}}=\frac{1}{8}(\bar{U}_{5}-2\tilde{\tilde{U}}_{5}+\tilde{U}_{5})[\rho_{A_{2}}^{f}]^{2}+\frac{1}{8}(\bar{U}_{4}-2\tilde{\tilde{U}}_{4}+\tilde{U}_{4})[\rho_{A_{2}}^{d}]^{2}\label{H_r_A2}
\end{align}
\begin{align}
H_{\rho_{B_{2}}}=\frac{1}{8}(\bar{U}_{5}+2\tilde{\tilde{U}}_{5}-\tilde{U}_{5})[\rho_{B_{2}}^{f}]^{2}+\frac{1}{8}(\bar{U}_{4}+2\tilde{\tilde{U}}_{4}-\tilde{U}_{4})[\rho_{B_{2}}^{d}]^{2}\label{H_r_B1}
\end{align}

We consider different channels separately, each time introducing order parameters into the Hamiltonian and summing up ladder series of vertex renormalizations.
 To simplify the formulas, below we set $m_h=m_e$.

\begin{figure}
\includegraphics[width=0.8\columnwidth]{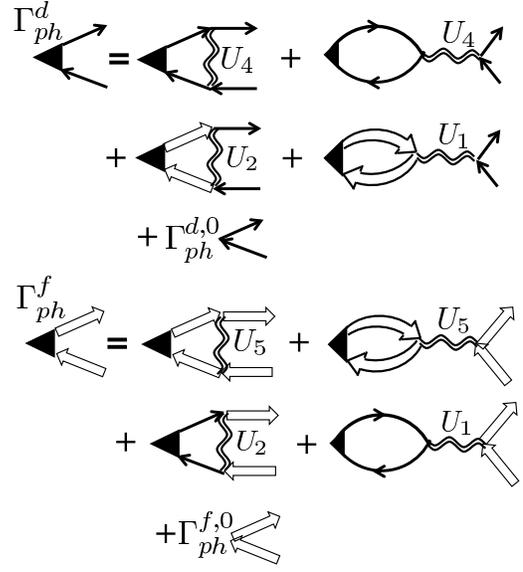}
\caption{
Diagrammatic representation of the Dyson equation for the interaction vertices in the Pomeranchuk channels
$A_{1}$ and $B_{1}$,
Eqs.~\eqref{chu_8} and \eqref{chu_31}.
The contributions from the interactions $\bar{U}_{1,2}$, $\tilde{U}_{4,5}$ and $\tilde{\tilde{U}}_{4,5}$,
which distinguish between $A_{1}$ and $B_{1}$ channels, are not shown.
\label{fig:Pom_vert}}
\end{figure}

\subsubsection{$B_{1}$ Pomeranchuk channel}

The $B_{1g}$ order parameter $\rho_{B_{1}}^{f(d)}$ changes sign under the $C_4$ rotation.
The vertices $\Gamma_{ph;B_{1}}^{f(d)}$  satisfy, see Fig.~\ref{fig:Pom_vert},
\begin{align}
\begin{bmatrix}\Gamma_{ph;B_{1}}^{d}\\
\Gamma_{ph;B_{1}}^{f}
\end{bmatrix}=M_{ph;B_{1}}\begin{bmatrix}\Gamma_{ph;B_{1}}^{d}\\
\Gamma_{ph;B_{1}}^{f}
\end{bmatrix}+\begin{bmatrix}\Gamma_{ph;B_{1}}^{d,0}\\
\Gamma_{ph;B_{1}}^{f,0}
\end{bmatrix}\,,\label{chu_8}
\end{align}
where $\Gamma_{ph;B_{1}}^{d(0)}$ and $\Gamma_{ph;B_{1}}^{f(0)}$ are the bare vertices  and
\begin{align}\label{M_ph}
M_{ph;B_{1}}=-2 \begin{bmatrix}u_{4}-2\tilde{u}_{4}+\tilde{\tilde{u}}_{4} & 2u_{1}-2\bar{u}_{1}-u_{2}+\bar{u}_{2}\\
2u_{1}-2\bar{u}_{1}-u_{2}+\bar{u}_{2} & u_{5}-2\tilde{u}_{5}+\tilde{\tilde{u}}_{5}
\end{bmatrix}\,.
\end{align}
Notice that there is no logarithm in the r.h.s. of (\ref{M_ph}). This is the consequence of the fact that particle-hole susceptibility at zero momentum transfer is non-logarithmical and is
 just the density of states at the Fermi level.

Using the values of the bare couplings from Eqs. ~\eqref{Hubbard_relation_1} and \eqref{dimensionless}, we obtain
\begin{align}
M_{ph;B_{1}}=12 \bar{g} \begin{bmatrix} 1& 1 \\
1& 1
\end{bmatrix}\,.\label{M_ph1}
\end{align}
Like before, there are two eigenvalues. One corresponds to the $d-$wave order parameter $n_{xz}-n_{yz}$ with the same sign on hole and electron pockets, for the other there is a
sign change  between $n_{xz}-n_{yz}$ on hole and electron pockets. By analogy with superconductivity, we label these order parameters as $d_{++}$ and $d_{+-}$.
  The eigenvalues of $M_{ph;B_{1}}$ are $\lambda_{P;++} =  24 \bar{g}$, $\lambda_{P;+-} = 0$. The coupling in the $d_{++}$ channel is attractive, but because there is no logarithm,
the $B_{1g}$ Pomeranchuk instability develops only if the coupling exceeds the critical value
\begin{align}
g > g_{ph;B_{1}} = \frac{\pi}{6 m}\, .
\label{n_n2}
\end{align}

\subsubsection{$A_{1}$ Pomeranchuk channel}

The order parameter $\rho_{A_{1}}^{f(d)}$ does not reduce the symmetry of the system and, as a result, the susceptibility in this channel never truly diverges.
Nevertheless, the $A_{1g}$  susceptibility can become large and, if the corresponding order parameter changes sign between the electron and hole pockets,
 the enhancement of the $A_{1g}$ susceptibility  leads to simultaneous shrinking (or enhancement) of electron and hole pockets.
The vertices $\Gamma_{ph; A_{1}}^{d,f}$  satisfy, see Fig.~\ref{fig:Pom_vert},
\begin{align}
\begin{bmatrix}\Gamma_{ph;A_{1}}^{d}\\
\Gamma_{ph;A_{1}}^{f}
\end{bmatrix}=M_{ph;A_{1}}\begin{bmatrix}\Gamma_{ph;A_{1}}^{d}\\
\Gamma_{ph;A_{1}}^{f}
\end{bmatrix}+ \begin{bmatrix}\Gamma_{ph;A_{1}}^{d,0}\\
\Gamma_{ph;A_{1}}^{f,0}
\end{bmatrix}\,,\label{chu_31}
\end{align}
where
\begin{align}\label{M_ph2}
M_{ph;A_{1}}=-2 \begin{bmatrix}u_{4}+2\tilde{u}_{4}-\tilde{\tilde{u}}_{4} & 2u_{1}+2\bar{u}_{1}-u_{2}-\bar{u}_{2}\\
2u_{1}+2\bar{u}_{1}-u_{2}-\bar{u}_{2} & u_{5}+2\tilde{u}_{5}-\tilde{\tilde{u}}_{5}
\end{bmatrix}\,.
\end{align}
The matrix $M_{ph;A_{1}}$ in Eq.~\eqref{M_ph2} differs from the matrix $M_{ph;B_{1}}$ in Eq.~\eqref{M_ph}  by signs in front of
$\tilde{u}_{4,5}$, $\tilde{\tilde{u}}_{4,5}$ and $\bar{u}_{1,2}$.
Substituting the bare values of the couplings from Eq.~\eqref{Hubbard_relation_1}, we obtain
\begin{align}\label{M_ph2_1}
M_{ph;A_{1}}=-4 \bar{g} \begin{bmatrix}1 & 1 \\
1 & 1
\end{bmatrix}\,.
\end{align}
We see that the matrix $M_{ph;A_{1}}$ has no positive eigenvalues.
As a result there is no  enhancement of the susceptibility in the $A_{1g}$ Pomeranchuk channel.

\subsubsection{$A_{2}$ and $B_{2}$ Pomeranchuk channels}

Equations \eqref{H_r_A2} and \eqref{H_r_B1} show that the interactions in the $A_{2g}$ channel is repulsive and the one in the $B_{2g}$ is attractive.
Analyzing the effects of the vertex renormalization in the same way as for other channels, we find that
the instability in the $B_2$ channel occurs at
{\begin{align}
 g_{ph;B_2} = \frac{m}{ 4\pi }.
\label{n_n2_1}
\end{align}
Comparing (\ref{n_n2}) and (\ref{n_n2_1}), we see that $g_{ph;B_2}> g_{ph;B_1}$. As a result, within RPA,
the instability in the $d$-wave Pomeranchuk channel occurs at a smaller coupling.

\section{RG analysis}
\label{sec:RG}

The existence of logarithmic renormalizations in both particle-hole and Cooper channels makes it necessary to study the coupling between the different channels.
Like we said in the Introduction, this can be  achieved by applying pRG technique. The pRG approach goes well beyond RPA and, in particular, includes non-ladder diagrams, which
 describe how fluctuations in one channel affect an effective interaction in the other channel.
pRG studies have been performed for pure band models with angle-independent interactions between band fermions~\cite{Chubukov2008,PhysicaC2009,Maiti2010} and, recently, for
orbitally-projected four pocket model with repulsive intra-orbital interactions~\cite{CKF2016}.  
To incorporate the Pomeranchuk channels, it is crucial to maintain the orbital content of the low-energy fermions.
Our model is the same as studied in Ref. ~\cite{CKF2016}, but some bare interactions are of different sign. We show that
  in our situation the system is in the basin of attraction of another fixed trajectory, and the system behavior is qualitatively different  from the one found in Ref. ~\cite{CKF2016}.

To simplify the presentation we again assume that $m_h=m_e$.  The derivation of pRG equations has been presented in Ref. \cite{CKF2016} and we use the results of that paper.

The pRG equations are split into three groups.
The two interactions ${\tilde{u}}_{5}$ and $\tilde{\tilde{u}}_{5}$ describe the subclass of scattering processes within the electron pockets.
 The flow of these two interactions decouple from that of other interactions and is only due to logarithmic renormalizations in
 the Cooper channel:
\begin{align}\label{u5_RG}
\dot{\tilde{u}}_{5} & =-(\tilde{u}_{5}^{2}+\tilde{\tilde{u}}_{5}^{2})\,,\notag\\
\dot{\tilde{\tilde{u}}}_{5} & =-2\tilde{u}_{5}\tilde{\tilde{u}}_{5}\, .
\end{align}
where the derivative is with respect to $L = \log{W/E}$, and $E$ is the running energy, at which the system is probed (all couplings vary with $L$).
 In our case the bare value ${\tilde {\tilde{u}}}_5 (L=0) =0$. Eq. (\ref{u5_RG}) shows that this coupling is then not generated under pRG.  The bare value of
  ${\tilde u}_5 (L=0)$ is $gm/(2\pi) >0$. According to (\ref{u5_RG}), this coupling then flows to zero under pRG.

Similarly, the two interactions involving fermions only near hole pockets also decouple and flow according to
\begin{align}\label{u4_RG}
\dot{\tilde{u}}_{4} & =-(\tilde{u}_{4}^{2}+\tilde{\tilde{u}}_{4}^{2})\,,\notag\\
\dot{\tilde{\tilde{u}}}_{4} & =-2\tilde{u}_{4}\tilde{\tilde{u}}_{4}\, .
\end{align}
Again, in our model the bare values are ${\tilde {\tilde{u}}}_4 (L=0)=0$, ${\tilde u}_4 (L=0)>0$.
According to (\ref{u4_RG}), ${\tilde {\tilde{u}}}_4$ is not generated, and ${\tilde u}_4 (L)$ flows to zero

The third group of pRG equations reads
\begin{align}
\dot{u}_{1} & =u_{1}^{2}+u_{3}^{2}\notag\label{RG_13_a}\\
\dot{\bar{u}}_{1} & =\bar{u}_{1}^{2}+\bar{u}_{3}^{2}\notag\\
\dot{u}_{2} & =2u_{1}u_{2}-2u_{2}^{2}\notag\\
\dot{\bar{u}}_{2} & =2\bar{u}_{1}\bar{u}_{2}-2\bar{u}_{2}^{2}\notag\\
\dot{u}_{3} & =-u_{3}u_{4}-\bar{u}_{3}\bar{u}_{4}+4u_{3}u_{1}-2u_{2}u_{3}-u_{5}u_{3}-\bar{u}_{5}\bar{u}_{3}\notag\\
\dot{\bar{u}}_{3} & =-\bar{u}_{3}u_{4}-u_{3}\bar{u}_{4}+4\bar{u}_{3}\bar{u}_{1}-2\bar{u}_{2}\bar{u}_{3}-u_{5}\bar{u}_{3}-\bar{u}_{5}u_{3}\notag\\
\dot{u}_{4} & =-u_{4}^{2}-\bar{u}_{4}^{2}-u_{3}^{2}-\bar{u}_{3}^{2}\notag\\
\dot{\bar{u}}_{4} & =-2u_{4}\bar{u}_{4}-2u_{3}\bar{u}_{3}\notag\\
\dot{u}_{5} & =-u_{5}^{2}-\bar{u}_{5}^{2}-u_{3}^{2}-\bar{u}_{3}^{2}\notag\\
\dot{\bar{u}}_{5} & =-2u_{5}\bar{u}_{5}-2u_{3}\bar{u}_{3}\, .
\end{align}
In our model  $\bar{u}_{i} (L=0) = 0$,  $i = 2,3,4,5$.
 Because the derivative $\dot{\bar {u}}_i$ is proportional to ${\bar u}_i$, the running ${\bar u}_i (L)$ simply remain zero:
\begin{align}\label{u_bar}
\bar{u}_i (L) = 0\, , \quad i=2-5\, .
\end{align}
With this simplification,  the equation for ${\bar{u}}_{1}$ also decouples from the rest and becomes
\begin{align}
\dot{\bar{u}}_{1} & =\bar{u}_{1}^{2}
\end{align}
Solving it we obtain
\begin{align}\label{RG_13_c}
\bar{u}_1(L) = \frac{ 1 }{ L' - L}\, ,
\end{align}
where $L' = [\bar{u}_1(L=0)]^{-1} = (2\bar{g})^{-1}$.

The remaining equations from the set (\ref{RG_13_a}) reduce to
\begin{align}\label{RG_13_b}
\dot{u}_{1} & =u_{1}^{2}+u_{3}^{2}\notag\\
\dot{u}_{2} & =2u_{1}u_{2}-2u_{2}^{2}\notag\\
\dot{u}_{3} & =-u_{3}u_{4}+4u_{3}u_{1}-2u_{2}u_{3}-u_{5}u_{3}\notag\\
\dot{u}_{4} & =-u_{4}^{2}-u_{3}^{2} \notag\\
\dot{u}_{5} & =-u_{5}^{2}-u_{3}^{2}\, .
\end{align}
The bare $u_4 (L=0) = u_5 (L=0)$. One can easily check that the running couplings remain equal, i.e., $u_4 (L) = u_5 (L)$.
The numerical solution of Eq.~\eqref{RG_13_b} is presented in Fig.~\ref{fig:RG}.
\begin{figure}
\includegraphics[width=0.8\columnwidth]{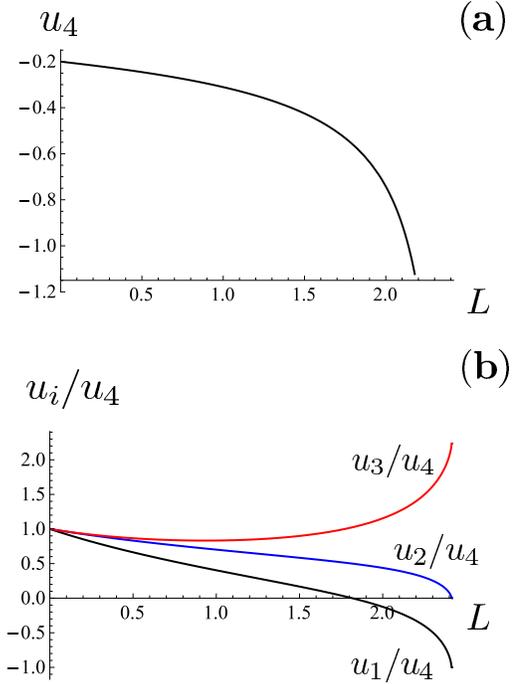}
\caption{The numerical solution of Eq.~\eqref{RG_13_b} for the RG flow of dimensionless vertices $u_i$ under the variation of the RG parameter $L = \log {W/E}$, where $W$ is of order bandwidth and $E$ is the running energy (or temperature).
 The initial condition is $u_i(L=0) = -2 \bar{g}$, $i = 1-5$, and we set $\bar{g} = 0.1$.
(a) The flow of the interaction $u_4$.  It remains negative, increases by magnitude, and diverges at the critical RG scale $L_0$. The divergence indicates that
 the system develops some form of order.
(b) The RG flow of the ratios of the interactions $u_i/u_4$.  As the RG scale $L$ approaches  the critical value $L_0$, the ratios tens to finite values, $u_1/u_4 = - \gamma_1^f = -1$, $u_2/u_4 = \gamma_2^f=0$, and $u_3/u_4 = \gamma_3^f=\sqrt{5}$,  in agreement with Eq.~\eqref{fixed}.\label{fig:RG}}
\end{figure}

Compared to the fixed trajectory found previously in Ref.~\cite{CKF2016},  the interactions $u_1$ and $u_4$ switch their respective roles.
In the model with purely repulsive interactions, considered in ~\cite{CKF2016}, inter-pocket density-density interaction $u_1>0$ gradually increases in the process of pRG flow, while intra-pocket interaction $u_4$ (initially positive) changes sign under pRG and gets more and more negative (attractive).
The pair hopping term $u_3$ is positive and, like $u_1$, it gradually increases under pRG.
 In our model, intra-pocket $u_4$ is negative from the start, and it just gets more negative in the pRG flow. At the same time, inter-pocket $u_1$ is initially negative, but it changes sign in the pRG flow and keeps increasing as a positive (repulsive) interaction.  The pair-hopping term $u_3$ is negative, and it gets more and more negative under pRG, much like $u_4$.
  As a result, in our case
  the interactions
     flow to a different fixed trajectory, than that found in Ref.~\cite{CKF2016}.

To analyse the fixed trajectories analytically, it is convenient to introduce  $v_i = - u_i$, $i=1-4$ and reduce the system of the remaining pRG equations to
\begin{align}\label{v}
\dot{v}_1 & = - v_1^2 - v_3^2 \notag \\
\dot{v}_2 & = - 2 v_1 v_2 + 2 v_2^2 \notag \\
\dot{v}_3 & =  2 v_3 v_4 + 2 v_3 v_2 - 4 v_3 v_1  \notag \\
\dot{v}_4 & = v_4^2 + v_3^2\,
\end{align}
with the initial conditions $v_i(L=0) = 2 \bar{g}$, $i = 1-4$.
We search for the fixed trajectory along which the ratios of the couplings tend to finite values.
Accordingly, we introduce
\begin{align}\label{gamma}
v_1 = - \gamma_1 v_4\, , \quad v_2 =  \gamma_2 v_4\, , \quad v_3 =  \gamma_3 v_4
\end{align}
where $\gamma_{i}$, $i = 1-3$ are constants.
Substitution of \eqref{gamma} in \eqref{v} yields the set of algebraic equations
\begin{align}\label{v_gamma}
\gamma_1 ( 1 + \gamma_3^2) & = \gamma_1^2 + \gamma_3^2 \notag \\
\gamma_2 ( 1 + \gamma_3^2) & = 2 \gamma_2 (\gamma_1 + \gamma_2 )\notag \\
\gamma_3 ( 1 + \gamma_3^2) & = 2 \gamma_3 ( 1 + \gamma_2 + 2 \gamma_1)\, .
\end{align}
The trivial fixed trajectory, $\gamma_1 = \gamma_2 = \gamma_3 = 0$ is unstable because the growth of $v_4$  makes  the solution with $v_3 = 0$ unstable,
 as follows from the third line in Eq.~\eqref{v}.
For the same reason the solutions $\gamma_1 =1, \gamma_2 = 0, \gamma_3 =0$ and $\gamma_1 =0, \gamma_2 = 1/2, \gamma_3 =0$
are unstable. One can also check that the solution $\gamma_1 = 1, \gamma_2 =-1/2,  \gamma_3 = 0$ is unstable and that there is no solution with
 $\gamma_1 =0, \gamma_2 = 0, \gamma_3 \neq 0$.

The remaining possibility  is that $\gamma_1 \neq 0$, $\gamma_3 \neq 0$ and $\gamma_2 =0$.
In this case the set of equations \eqref{v_gamma} reduces to two equations,
\begin{align}\label{e13}
\gamma_1 ( 1 + \gamma_3^2) & = \gamma_1^2 + \gamma_3^2 \notag \\
 ( 1 + \gamma_3^2) & = 2 ( 1 + 2 \gamma_1)\, .
\end{align}
It follows from the second line of Eqs.~\eqref{e13} that $\gamma_3^2 = 1 + 4 \gamma_1$.
The first line of Eqs.~\eqref{e13} can be written as
\begin{align}\label{v_gamma1}
\gamma_3^2( \gamma_1 - 1) = \gamma_1(\gamma_1 - 1)\, .
\end{align}
Equation \eqref{v_gamma1} offers two alternatives.
The first is $\gamma_1 = 1$, and the second is $\gamma_1 = \gamma_3^2$.
The latter possibility is however not viable as in combination with the second line of Eq.~\eqref{e13} it results in the relation
$3 \gamma_3^2  + 1 = 0$ that cannot be satisfied.
We therefore have $\gamma_1 = 1$, and from the second line of Eq.~\eqref{e13}, $\gamma_3 = \pm \sqrt{5}$.
To fix the sign of $\gamma_3$ we note that the unstable fixed trajectory $\gamma_1 = 1$, $\gamma_2 = 0$, $\gamma_3 = 0$ is the separatrix that cannot be crossed under the pRG flow.
In other words the interaction $u_3$ maintains its sign under pRG, i.e., it is fixed by the initial conditions.
Since $v_3(L=0) >0$,  the fixed trajectory is
\begin{align}\label{fixed}
\gamma_1^f = 1, \gamma_2^f = 0, \gamma_3^f = \sqrt{5}\, .
\end{align}
Let us verify that the fixed trajectory set by Eq. \eqref{fixed} is stable.
 For this we allow  the coefficients $\gamma_i$, $i = 1-3$ to vary slightly, rewrite the set of pRG equations as the set for $\gamma_i (L)$
\begin{align}\label{stab_1}
\dot{\gamma}_1 &= v_4 [ (\gamma_1^2 + \gamma_3^2) - \gamma_1 ( 1 + \gamma_3^2) ]
\notag \\
\dot{\gamma}_2 &= v_4 [ 2 \gamma_1 \gamma_2 + 2 \gamma_2^2 - \gamma_2 ( 1 + \gamma_3^2) ]
\notag \\
\dot{\gamma}_3 &= v_4 [ 2 \gamma_3 + 2 \gamma_2 \gamma_3 + 4 \gamma_3 \gamma_1 - \gamma_3 ( 1 + \gamma_3^2) ]\, .
\end{align}
and linearize Eqs.~\eqref{stab_1} in small deviations, $\delta \gamma_i = \gamma_i -  \gamma_i^f$.
 The set of linear differential equations can be cast into the matrix form
\begin{align}
\delta  \dot{\gamma}_i =  \sum_{j= 1}^3 \Lambda_{ij} \delta \gamma_j\, ,
\end{align}
with
\begin{align}
\Lambda = -v_4 \begin{bmatrix}
4 & 0 & 0 \\
0 & 4 & 0 \\
4 \sqrt{5} & 2 \sqrt{5} & 10
\end{bmatrix}\, .
\end{align}
We see that $\Lambda$ is  negative definite.
As a result the fixed trajectory defined by Eq.~\eqref{fixed} is stable.

Along the fixed trajectory set by Eq.~\eqref{fixed}  the fourth equation from Eq.~\eqref{v} becomes $\dot{v}_4  = 6 v_4^2$.  Assuming that this equation is valid
starting already from small $L$, we find the solution in the form
\begin{align}\label{v_4}
v_4(L) = \frac{ v_4(0) }{ 1 - 6 L v_4(0)} = \frac{ 1  }{ 6(L_0 - L)}\, ,
\end{align}
where
\begin{align}
L_0 = \frac{1}{6 v_4 (0)}.
\end{align}
The initial value $v_4(0) = - u_4 (0) = 2\bar{g}$. Hence
\begin{align}
L_0 =
\frac{1}{12 \bar{g}} \, .
\end{align}
Comparing with Eq. (\ref{RG_13_c}) we see that $L_0 < L'$, hence the couplings $v_i$ (and $u_i = -v_i$) diverge at a smaller $L$ (i.e., larger energy) than
${\bar u}_1$.  Then,   ${\bar u}_1$ can be neglected compared to $u_i$ near the fixed trajectory.

Summarizing the pRG analysis, we find that for our model there exists one stable fixed trajectory along which
\begin{align}\label{RG_Fixed}
u_1(L) &= \frac{ 1  }{ 6(L_0 - L)}\, ,
u_3(L) = -\frac{\sqrt{5} }{ 6(L_0 -  L)}\, ,
\notag \\
u_4(L) & =u_5(L)=- \frac{ 1  }{ 6(L_0 - L)}\, ,
\end{align}
and rest of the interactions are either zero, or flow to zero, or increase but at a smaller rate than the interactions listed in Eq. (\ref{RG_Fixed}).

\section{Hierarchy of instabilities within pRG}
\label{sec:fixed}

We now  reexamine the hierarchy of instabilities using the renormalized, scale-dependent interactions, listed in  Eq.~\eqref{RG_Fixed}.
For this we follow Ref. \cite{CKF2016} and earlier functional RG works (Refs. \cite{Platt2009}) and obtain and solve the RG equations for the vertices $\Gamma_i$ in different channels, using the running
 couplings as inputs.  We then  use the running vertices to compute the susceptibilities in different channels, and compare the exponents for the susceptibilities $\chi_j \propto
  1/(L_0-L)^\alpha_j$, where $j$ labels different channels.
  Like in other RG-based approaches, we assume that the instability at $L = L_0$ will lead to the development of a non-zero mean value of the order parameter,
  for which $\alpha_j$ is the largest.   We will not present the details
 of the derivation of RG equations as the computational steps have been already described in Ref. \cite{CKF2016}.  We, however, discuss the computations of the running susceptibility
  in the Pomeranchuk channels in some more detail.

\subsection{Magnetism}

\begin{figure}
\includegraphics[width=0.6\columnwidth]{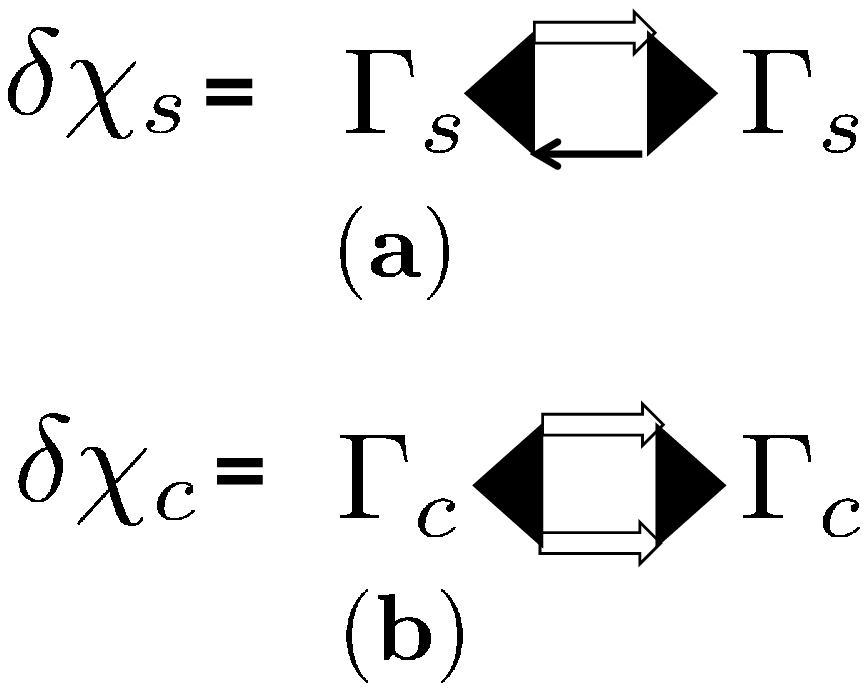}
\caption{
Diagrammatic representation of pRG expressions for the susceptibility in SDW channel (a) and Cooper channel (b) that give rise to Eqs.~\eqref{G_SDW_2} and
\eqref{G_SDW_2_1}, respectively.
\label{fig:RG_suscept}}
\end{figure}

Within RPA, the intraorbital SDW does not develop, while interorbital SDW develops at a lower $T$ than superconductivity and CDW order.
   The result of pRG analysis is somewhat different. Namely,
  real intra-orbital order ${\bm{s}}_{1,2}^{r}$ does not develop because the coupling  $- u_1 - u_3$ remains positive (repulsive) under pRG.  But
  for imaginary  intra-orbital order ${\bm{s}}_{1,2}^{i}$  the corresponding dimensionless coupling $-u_1+u_3$ becomes positive and grows in the process of RG flow.
The RG equation for the vertex function $\Gamma_{s}^i$ (introduced in the same was as in Sec. III) is
\begin{align}\label{G_SDW}
\frac{d \Gamma_{s}^i }{ d L } =(u_1 - u_3) \Gamma_{s}^i \, .
\end{align}
where $u_1 = u_1 (L)$ and $u_3 = u_3 (L)$ are the running couplings. The boundary condition is  $\Gamma_{s}^i (L=0) = \Gamma_{s}^i(0)$.
The solution of Eq.~\eqref{G_SDW} along the fixed trajectory, i.e., with $u_1(L)$ and $u_3 (L)$ given by Eq.~\eqref{RG_Fixed}, is
\begin{align}\label{G_SDW_1}
\Gamma_{s}^i = \frac{\Gamma_{s}^i (0)}{ (L_0 - L)^{\beta_{s}^i}}\, , \quad  \beta_{s}^i = \frac{ 1 + \sqrt{5}}{6}\,.
\end{align}
The  running susceptibility $\chi^i_{s} (L)$ evolves according to
\begin{align}\label{G_SDW_2}
\frac{d \chi_{s}^i }{ d L } =[ \Gamma_{s}^i ]^2 \, ,
\end{align}
see Fig.~\ref{fig:RG_suscept}a.
Substituting  $\Gamma_{s}^i (L)$ from Eq.~\eqref{G_SDW_1} and integrating over $L$, we obtain
\begin{align}
\chi_{s}^i & \propto (L_0 - L)^{- \alpha_{s}^i}\, ,
\notag \\
\alpha_{s}^i &= 2 \beta_{s}^i - 1 = \frac{ \sqrt{5} - 2}{3} \approx 0.08\, .
\end{align}

The interactions in inter-orbital SDW channels with real and imaginary order parameters are attractive already at the bare level, and keep increasing under pRG.
The behavior of the corresponding  ${\bar \Gamma}_{s}^{r,i}$ is governed by the running ${\bar u}_1$. The latter diverges, but at $L=L'$, which is larger than $L_0$.
As a result, the instability in the intra-orbital SDW channel occurs at higher running energy, and, hence, at a higher temperature.

\subsection{Superconductivity}

We now consider  susceptibilities in the  superconducting channels.
First, $A_{1}$ and $B_1$ channels remain degenerate because the running couplings in these two channels differ by ${\bar u}_j$, $j=3,4,5$ (see Eqs. ~\eqref{s_SC} and
 \eqref{d_SC}). These couplings are zero at the bare level and
remain zero under pRG, see Eq.~\eqref{u_bar}.  The interaction in $s^{+-}$ and $d^{+-}$ channels is $u_4 - u_3$. This interaction is repulsive along the fixed trajectory, hence the
corresponding susceptibility does not diverge. The interaction in $s^{+-}$ and $d^{+-}$ channels is $u_4 + u_3$, and this one is negative (attractive) along the fixed trajectory.
The RG equation for the SC vertex in $s^{++}$ and $d^{++}$ channels is
\begin{align}\label{SC_f}
\frac{d \Gamma^{s,d}_{sc}}{ d L } =-(u_4 + u_3) \Gamma^{s,d}_{sc }\, ,
\end{align}
Solving this equation we find
\begin{align}\label{G_SDW_1_1}
\Gamma^{s,d}_{sc} = \frac{\Gamma^{s,d}_{sc} (0)}{ (L_0 - L)^{\beta^{s,d}_{sc}}}\, , \quad  \beta^s_{sc} = \beta^d_{sc} = \frac{ 1 + \sqrt{5}}{6}\,.
\end{align}
The  running susceptibilities $\chi^{s,d}_{sc} (L)$ again evolve according to
\begin{align}\label{G_SDW_2_1}
\frac{d \chi^{s,d}_{sc} }{ d L } =[ \Gamma^{s,d}_{sc} ]^2 \, ,
\end{align}
see Fig.~\ref{fig:RG_suscept}b.
Substituting  $\Gamma^{s,d}_{SC} (L)$ from Eq.~\eqref{G_SDW_1_1} and integrating over $L$, we obtain
\begin{align}
\chi^{s,d}_{sc} & \propto (L_0 - L)^{- \alpha^{s,d}_{sc}}\, ,
\notag \\
\alpha^{s,d}_{sc} &= 2 \beta^{s,d}_{sc} - 1 = \frac{ \sqrt{5} - 2}{3} = \alpha^i_{sc} .
\end{align}
We see that the susceptibilities in $s^{++}$ and  $d^{++}$ channels have the same exponents as the susceptibility in intra-orbital SDW channel with imaginary order parameter.

For $B_{2}$ channel, the tendency towards pairing is suppressed at low energies because $\tilde{u}_{4,5}$ and $\tilde{\tilde{u}}_{4,5}$ flow to zero.

\subsection{CDW order}
The same analysis as in the previous two subsections shows that the susceptibility for real intra-orbital order parameter
 $\delta_{1,2}^{r}$ diverges as $L$ approaches $L_0$, while the susceptibilities in other CDW channels do not diverge.
The divergent CDW susceptibility scales as
\begin{align}\label{CDW_1}
\chi_{c}^r \propto (L_0 - L)^{- \alpha_{c}^r}\, .
\end{align}
 where $\alpha_{c}^r = (\sqrt{5} - 2)/3$.  This exponent is the same as $\alpha^{s,d}_{sc}$ and $\alpha^i_{s}$, i.e., within RG the
 susceptibilities in all these channels scale with each other.

The susceptibilities in the inter-orbital CDW channels
 remain regular, i.e., the corresponding order parameters do not develop at $L=L_0$.

\subsection{Pomeranchuk order}

Within RPA, the instability in any of Pomeranchuk channels develops only when the interaction exceeds a certain threshold.  This is
the consequence of the fact that the renormalization of the Pomeranchuk vertex is determined by the convolution of the two fermion propagators at vanishing transferred
 momentum and zero transferred frequency.
  This convolution gives a constant
 (equal to the density of states at the Fermi level),  but not a logarithm.

Within pRG, we need to evaluate the vertex at a running frequency.   The triple vertex, shown in Fig.~\ref{fig:Pom_RG}, depends on two external frequencies, $E$ and $E''$ (the third one is $E''+E$ by frequency conservation).  To obtain susceptibility, we will need o integrate over $E''$.
We assume and then verify that relevant $E''$ are comparable to $E$.

 If we re-evaluate the convolution of the two propagators at a finite $E$ and $Q=0$, we  obtain that the result vanishes, because the
 poles in the two fermionic propagators are in the same half-plane of a complex frequency.  Does this imply that Pomeranchuk vertex is not renormalized within RG?  We argue that
  it doesn't, and the Pomeranchuk vertex does flow under RG. The reason is that to obtain  vertex renormalization we actually need to compute  the product of the two fermionic propagators and the interaction. This combination is expressed via the convolution of the two fermionic propagators at a finite $E$ and $Q=0$ only if the interaction is static.   But the running interaction is not a constant but rather a function of the running fermionic frequency $E'$ and also of external $E'' \sim E$.
   As the consequence, when we compute the renormalization of the
  Pomeranchuk vertex at a given energy $E$,  we need to evaluate the momentum and frequency integral of the product of the two propagators and the running interaction
  (see Fig.~\ref{fig:Pom_RG}):
  \begin{equation}
  I (E) = \int d^2k d E' \frac{1}{i E' - \epsilon_k} \frac{1}{i(E' + E) - \epsilon_k}~  U_j (E,E')
  \label{chu_1}
  \end{equation}
   where $U_j$ is one of the interactions (see Fig.~\ref{fig:Pom_RG}).
    One can verify that, to logarithmic accuracy, the dependence of the interaction $U_j (E,E')$ on $|E|$ and $|E'|$ can be cast as the dependence on $L = \log {W/(|E| + |E'|)}$.
\begin{figure}
\includegraphics[width=0.8\columnwidth]{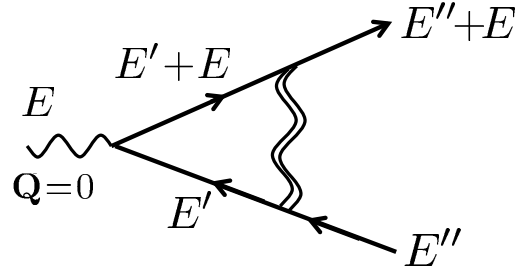}
\caption{The diagrammatic representation of the lowest-order vertex  renormalization  in the Pomeranchuk channel.
Double wavy line represents the running interaction $U_i (L)$.
The external $E$ and $E'' \sim E$ can be regarded either as  frequencies in $T=0$ calculations or as a temperature.  In the first case the integral over internal $E'$  does not vanish
 because the running interaction is also a function of $E'$, and equals to the density of states $N_F$ times the coupling at a scale $E$.
    In the second case,  the interaction is treated as static, but  the
 convolution of the two fermion propagators is again nonzero and equal to the density of states $N_F$.
 \label{fig:Pom_RG}}
\end{figure}
    Because  $U_j (E, E')$, has a non-analytic dependence on the running $E'$,
      the integrand in (\ref{chu_1}) contains branch cuts in addition to the poles,
       and the branch cuts
     are present in both  half-planes of complex $E'$.  In this situation, in is more convenient to first evaluate the integral over $d^2k$ and then over $dE'$.
       For this, we  subtract from $U_j (E, E')$ its constant
      value at $E, E'= W$. This does not change $I(E)$ because, as we just said,  the term we
      subtract gives zero contribution to $I(E)$.
         The integrand in (\ref{chu_1}) with $U_j (E, E') - U_j (W)$ converges and the integration can be done in any order.
           Taking for definiteness fermions  near an electron pocket, transforming from the integration over $d^2 k$ to integration
          over $d \epsilon_k$ via $\int d^2k = N_F \int^W_{-E_F} d \epsilon_k$,
           and integrating over $\epsilon_k$ first,  we obtain for positive $E > E_F$,
        \begin{equation}
  I (E) \sim  N_F \int_{E_F}^E  \frac{d E'}{E} \left(U_j (E,E') - U_j (W)\right)
   \label{chu_2}
  \end{equation}
  or, in logarithmical variables
  \begin{equation}
  I(L)  \sim N_F\int_{L-\log{2}}^{L} e^{L-L'}  \left(U_j (L') - U_j (W)\right) \sim  u_j(L)
    \label{chu_3}
  \end{equation}
    Evaluating this integral to logarithmical accuracy, we find that  one loop renormalization of the Pomeranchuk vertex
  $\Gamma_{ph} (L)$ yields $\Gamma_{ph} (L) \propto u_j (L)$, i.e., the vertex at a scale $L$ is proportional to the running interaction at the same scale $L$.

Alternatively, we can view the RG energy variable $E$ as a temperature  and consider how the couplings vary as one progressively integrate out
 fluctuations at a higher $T$.  In this approach the susceptibilities in all channels  are the static ones ($E=0$), but taken at a finite $T$.
 The integration over $E'$ in (\ref{chu_1}) now has to be replaced by the summation over Matsubara frequencies.  A static interaction can be taken outside
  the frequency summation, but the latter now gives a finite result because regularization by a finite $T$ yields the same result --
   the density of states $N_F$ -  as the evaluation of the convolution of the two $G's$  at $T=0$, $E=0$ and $Q \to 0$.  Furthermore,
     relevant internal $E'$ of order $T$.  Hence, the vertex at a given $T$ is proportional to the interaction at the same $T$, i.e., in logarithmical variables we have
    the same dependence $\Gamma_{ph} (L) \propto u_j (L)$ as in $T=0$ analysis with frequency $E$ as the running variable.

  Another consequence of the pRG flow of the couplings that the interplay between the running interactions $u_j (L)$ is different from the one between the bare interactions, chiefly because
  $u_1 (L)$ changes its sign in the process of RG flow and becomes positive, i.e., attractive in $A_{1g}$ and $B_{1g}$ Pomeranchuk channels.

 \begin{figure}
\includegraphics[width=0.8\columnwidth]{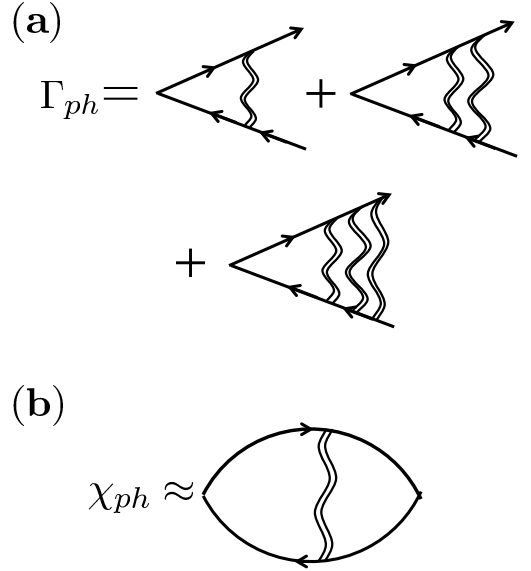}
\caption{
(a) Series of ladder diagrams for the vertex function in the Pomeranchuk channel.
 Compared to the diagram in Fig.~\ref{fig:Pom_RG}, these diagrams account for the shift of the critical $L$ from $L-0$ to $L_{ph} = L_0 -1$.
 This shift is beyond logarithmic accuracy and we neglect it when compare Pomeranchuk and other channels.
(b) The  contribution to the Pomeranchuk susceptibility to first order in the running coupling.
The logarithmic enhancement of the Pomeranchuk susceptibility is due to $1/(L_0-L)$ scaling of the interaction at the running pRG scale $L$.
\label{fig:Pom_RG2}}
\end{figure}

We now sum up ladder series of renormalizations of $\Gamma_{ph}$ (see Fig.~\ref{fig:Pom_RG2}a).
 These are the same series as we summed up for SDW and SC vertices.  The summation leads to the same matrix equations for the full
 vertices $\Gamma^{f,d}_{ph;B_1} (L)$ as in  the RPA analysis of Pomeranchuk instabilities, Eq. \eqref{M_ph},
  but  now  $u_i$ are the running interactions.
  Along the fixed trajectory we obtain
\begin{align}\label{M_ph_2}
M_{ph;B_{1}}=-2 \begin{bmatrix} u_{4} & 2u_{1}\\
2u_{1} & u_{5}
\end{bmatrix}\, .
\end{align}
Substituting $u_j (L)$ from  Eq.~\eqref{RG_Fixed} we  re-express (\ref{M_ph_2}) as
\begin{align}\label{M_ph_3}
M_{ph;B_{1}}
=
\frac{1}{3(L_0 - L)} \begin{bmatrix} 1 & - 2\\
- 2 & 1
\end{bmatrix} \,.
\end{align}
The two eigenvalues of this  matrix are
\begin{align}\label{M_ph_4}
\lambda_{B_1, ++} = -\frac{ 1 }{ 3 (L_0 -L)} \, , \quad \lambda_{B1, +-} = \frac{ 1 }{ (L_0 -L)}\,
\end{align}
 We remind that notations $++$ and $+-$  refer to $B_1$ (d-wave) order parameters $n_{xz}-n_{yz}$
  with the same (opposite) sign on hole and electron pockets.
It follows from Eq.~\eqref{M_ph_4} that the Pomeranchuk instability is $d^{+-}$ channel.  This is different from RPA, where we found the leading instability in
 $d^{++}$ channel.  The discrepancy with RPA is the consequence of the sign reversal of the interaction $u_1 (L)$ in the process of pRG flow.

The instability towards $d^{+-}$ nematic order occurs when $\lambda_{ph;+-} =1$ i.e., at
$L_{ph} = L_0 -1$.  This difference, however,  is beyond logarithmical accuracy and we neglect it, i.e., approximate $L_{ph}$ by $L_0$.
 In the diagrammatic approach, this corresponds to keeping only the leading term in the ladder series for $\Gamma^{f,d}_{ph,B_1} (L)$, see Fig.~\ref{fig:Pom_RG2}b).

 More important is the fact that near the instability the Pomeranchuk vertex scales as
\begin{align}
\Gamma^{f,d}_{ph; B_1} (L)  \propto \frac{1}{L_{0} -L}
\end{align}
 i.e., the corresponding $\beta_{ph} =1$, while for other channels  $\beta$ is close to $1/2$.

Using the same reasoning in the computation of the  susceptibility in $B_{1}$ Pomeranchuk channel, we find
\begin{align}
\chi_{ph; B_1} (L)  \propto~\frac{1}{L_{0} -L}
\end{align}
i.e., the exponent for the Pomeranchuk susceptibility in the $B_1$ channel is $\alpha_{ph} =1$, much larger than $\alpha =0.08$ in SDW, CDW, and $s$and $d$-SC channels.
This difference in the numbers is important because compared to other susceptibilities the one in the   Pomeranchuk channel contains additional factor of a running coupling
$u(L)$  due to the absence
of the logarithm in the vertex renormalization. At some distance from $L = L_0$, $u(L) \sim 1/L_0$ is small, hence $\chi_{ph;B_1}$ is parametrically smaller than other susceptibilities.
  If the exponents in the Pomeranchuk and other channels were similar in magnitude, $\chi_{ph, B_1}$ would exceed susceptibilities in
other channels only at $L$ near $L_0$, where $u(L) \geq 1$ and the accuracy of one-loop pRG is questionable.  Because all other $\alpha$ are small and $\alpha_{ph, B_1} =1$,
 $\chi_{ph;B_1}$ becomes larger than the susceptibilities in other channels at much larger distance from $L_0$, when one-loop pRG is likely still valid.

We refrain from discussing the susceptibility in $A_1$ Pomeranchuk channel because, as we said, this susceptibility  does not actually diverge.
The interactions in $A_2$ and $B_2$ Pomeranchuk channels flow to zero under pRG, i.e.,  the corresponding susceptibilities do not diverge.

We note that the Pomeranchuk order changes the shape of the Fermi surface, but leaves fermionic excitations gapless. This leaves the possibility that superconductivity
 emerges at a lower temperature inside the nematic phase, as it happens in FeSe.
In our model, the behavior in the nematic phase may be even more complex as the susceptibilities in SC, SDW, and CDW  channels are expected to continue to grow  below the nematic transition. These three channels compete for the secondary instability, and the outcome of this competition depends  on the details of the electronic structure,
such as the degree of nesting between electron and hole pockets and the ratio of hole and electron masses.
The detailed study of this competition is beyond the scope of this work.

\section{Discussion}
\label{sec:Discussion}

In this paper  we performed a detailed study of potential two-fermion instabilities in a model of FeSCs with the interaction tailored to favor $C_4$-breaking orbital order.
In distinction to the two-orbital model with the same interaction considered in earlier works,  we used the correct four-pocket band structure
 with two hole pockets at the centre of the BZ and two electron pockets at its boundaries.  We kept the
 orbital content of low-energy excitations, what allowed us to include orbital fluctuations along with SDW, CDW, and SC fluctuations.
 We have shown that the interplay between different interaction channels substantially affects the hierarchy of the ordering tendencies.

We first analysed the model within RPA which neglects the interplay between different channels.
 We found that the highest- $T$ instabilities at weak coupling are in $s^{++}$ and $d^{++}$ SC channels ($s-$wave and $d-$wave with no sign change of the gap
 between hole and electron pockets), and in an intra-orbital CDW channel with transferred momenta $(0,\pi)$ or $(\pi,0)$.  The instability temperature is the same in all three channels.
 The sign-preserving SC state wins over sign changing states ($s^{+-}$ and $d^{+-}$ because in our model intra-orbital interaction is attractive.
 The degeneracy between $s^{++}$ and $d^{++}$ channels is the  consequence of the absence of the Hund  coupling $J'$ which would give rise to the tunnelling of Cooper pairs
  of electrons on $d_{xz}$ orbitals into Cooper pairs on $d_{yz}$ orbitals and vice versa.

 There is also attractive interaction in inter-orbital SDW and CDW channels. The instability temperature is the same in both channels, but it is lower than that in the three
 leading channels.  In addition, there is attractive interaction in $B_{1g}$, $A_{2g}$ and $B_{2g}$  Pomeranchuk channels, but the instability there occurs only when the coupling
  exceeds a certain threshold. The threshold value is the smallest in $B_{1g}$ channel.

We next studied the effect of the coupling between different channels.  We applied RG technique, obtained and  solved the set of parquet RG equations for the interactions, and
 identified the stable fixed trajectory as the asymptotic solution of these equations.
On a fixed trajectory the ratios of any two interactions is just a number.
We found that the  fixed trajectory in our model is notably distinct from the one obtained for the model with intra-orbital and inter-orbital repulsion.
In the latter case
 the intra-pocket interactions flip the sign before the system reaches the fixed trajectory.
  This  turns intra-pocket repulsion into an attraction. The interaction describing the
inter-pocket tunnelling of Cooper pairs
remains attractive and becomes the strongest under pRG.  This gives rise to $s^{+-}$ superconductivity. The interplay between different couplings is such that
SC wins over intra-orbital SDW, but the SC susceptibility gets weakened by the competition and may loose to   $d-$wave Pomeranchuk order.

In the model which we considered here,
 intra-pocket interactions and the inter-pocket Cooper pair tunnelling are attractive at the bare level and remain attractive in the process of pRG flow, while
 inter-pocket density-density interaction flips sign under pRG from attraction to repulsion.
 As the consequence, four channels are degenerate along the fixed trajectory in the sense that the corresponding susceptibilities all diverge at the same energy (temperature) and
 with the same exponent. These four are $s^{++}$ and $d^{++}$ SC channels, intra-orbital CDW channel and intra-orbital SDW channel, all with real order parameter.
  Due to strong competition between that many channels, the exponent for the susceptibilities is quite small, $\alpha =0.08$, i.e., the four susceptibilities barely diverge at the critical RG scale.   Meanwhile, $d-$wave Pomeranchuk channel (the one with the $C_4$-breaking orbital order parameter $n_{xz} - n_{yz}$)
  remains attractive during pRG flow, and the exponent for the $d-$wave Pomeranchuk susceptibility is $\alpha=1$.
 At  intermediate RG scales, Pomeranchuk susceptibility is smaller than the ones in four other singular channels because of the absence of a logarithm in
 the particle-hole
 polarization bubble at zero momentum transfer.  But near  the critical RG scale $L =L_0$, Pomeranchuk susceptibility is the largest because of larger exponent
 $\alpha$. Because of large
 numerical difference between $\alpha =1$ in the $d-$wave Pomeranchuk channel and $\alpha =0.08$ in other channels, the susceptibility in the Pomeranchuk channels
  becomes the largest already at a substantial  distance from the critical RG scale $L_0$, when one-loop pRG approach is  under control in the sense that two-loop corrections
  are numerically small.  The outcome is that in the model that we considered in this work the leading candidate for the instability already
  at weak coupling is a spontaneous orbital order.
  The verification of this result in numerical studies is called for.\\

From physics perspective, the microscopic mechanism for the Pomeranchuk order in the model of Eq. (5) is two-fold.  First,  growing CDW fluctuations not only boost the attraction in the current (imaginary) intra-orbital SDW channel and in s+- and d-wave superconducting channels, but also boost attractive interaction in the d-wave Pomeranchuk channel  Second,  SC and SDW channels compete with CDW channel,  and this competition reduces SDW and superconducting susceptibilities.   d-wave Pomeranchuk channel does not compete with other channels,  and the  susceptibility  in this channel is not reduced.  This is  why the exponent in this channel is larger than those in the other channels.
 We also emphasize that pRG analysis goes beyond RPA. In RPA, there is an  instability in the d-wave Pomeranchuk channel in the model of Eq. (5), but it holds only  when $g$ exceeds the critical value $g_c$, and is always secondary to superconductivity.  The pRG analysis includes two effects not present in RPA: (i) the boost of the interaction in the Pomeranchuk channel by CDW fluctuations, and (ii) the reduction of the susceptibility in the superconducting channel due to competition with CDW.

An obvious issue is how sensitive are our results to the modification of the Hamiltonian, particularly the modification of the interaction in Eq. (5), and
of the degree of nesting and the value of the chemical potential in the electronic structure, and of the strength of the interaction.
On this, we make a couple of observations.
 First, in pRG approach, nesting (by which mean near equal size of hole and electron pockets) does not play the crucial role.  All what
 matters for pRG is the  opposite sign of the dispersion of excitations near hole and electron pockets.   Second, the pRG flow holds at energies between the bandwidth and the Fermi energy and as such is not sensitive to the details of the electronic structure at energies smaller than $E_F$. In this respect, variations of the chemical potential over energy range smaller than the Fermi  energy will not affect the pRG flow.  The variation of the ratio of hole and electron masses also does not affect the behavior of the couplings along the fixed RG trajectory and the hierarchy of instabilities. Third, in any one-loop RG-based  study there are two  assumptions: (i)  that the channel for the leading instability gets selected   already within the applicability range of RG (i.e., at energies above $E_F$), and (ii) that the terms beyond one-loop RG do not affect this selection.     The larger is the bare coupling ($g$ in Eq. (5)), the more important are the terms beyond one-loop RG.   Neither we nor other groups analysing the RG flow in multi-orbital systems  went beyond one-loop RG simply because one-loop RG equations  are already complex enough.  Whether the RG results remain valid at $g$ comparable to the bandwidth should be addressed by comparing RG phase diagram with the results of numerical studies.
 We also note that the huge difference between the exponents in the Pomeranchuk channel and in other channels in our model is a guarantee that Pomeranchuk channel wins even in a more complex model, where SC, SDW, and CDW susceptibilities become non-equal, and  one of the corresponding exponents become larger. Indeed, this holds only as long as all exponents remain substantially smaller than one.  If this is not the case, our  reasoning breaks down.

A more subtle aspect, which is not fully understood at the moment,
 is  whether the fact that  CDW, SDW, and SC orders all may potentially break $C_4$ symmetry plays the role in the system's selection of the
 $C_4$ breaking Pomeranchuk order as the leading instability.  Indeed, stripe CDW and SDW break $C_4$, and the degeneracy between $s$ and $d$-wave SC orders opens the way to
  $s+id$ state, which also breaks $C_4$. At the same time, whether or not CDW or SDW order is a  stripe or a checkerboard can be determined only by analyzing the interplay between 4-th order terms in SDW and CDW order parameters. Such terms are of eighth order in fermions and are beyond one-loop RG.

\begin{acknowledgements}
We thank G. Blumberg, L. Classen, R.M. Fernandes, V. K. Thorsm\o lle, O. Vafek  and R. Xing for useful discussions.
MK acknowledges the support by the Israel Science Foundation, Grant No. 1287/15 and NSF DMR-1506668.
AVC acknowledges the support by  the Office of Basic Energy Sciences, U.S. Department of Energy, under award DE-SC0014402.
\end{acknowledgements}

\end{document}